\documentclass[useAMS,usenatbib]{mn2e}

\usepackage{graphicx}
\usepackage{epsfig}

\usepackage{amsmath}  
\usepackage{tabularx}  
\usepackage{multirow}

\newcommand{\ha}{H$\alpha$}

\def\vhel{\ifmmode{V_{{\rm HEL}}}\else{$V_{{\rm HEL}}$}\fi}
\def\vsys{\ifmmode{V_{\rm sys}}\else{$V_{\rm sys}$}\fi}
\def\kms{\ifmmode{~{\rm km\,s}^{-1}}\else{~km~s$^{-1}$}\fi}
 \def\vlsr{\ifmmode{v_{\rm lsr}}\else{$v_{\rm lsr}$}\fi}

\def\ltsim{\ifmmode\stackrel{<}{_{\sim}}\else$\stackrel{<}{_{\sim}}$\fi}
\def\gtsim{\ifmmode\stackrel{>}{_{\sim}}\else$\stackrel{>}{_{\sim}}$\fi}

\newcommand{\ie}{\textit{i.e. }}

\def\reff@jnl#1{{\rm#1\/}}

\def\aj{\reff@jnl{AJ}}                  
\def\araa{\reff@jnl{ARA\&A}}            
\def\apj{\reff@jnl{ApJ}}                
\def\apjl{\reff@jnl{ApJ}}               
\def\apjs{\reff@jnl{ApJS}}              
\def\ao{\reff@jnl{Appl.Optics}}         
\def\apss{\reff@jnl{Ap\&SS}}            
\def\aap{\reff@jnl{A\&A}}               
\def\aapr{\reff@jnl{A\&A~Rev.}}         
\def\aaps{\reff@jnl{A\&AS}}             
\def\azh{\reff@jnl{AZh}}                        
\def\baas{\reff@jnl{BAAS}}              
\def\jrasc{\reff@jnl{JRASC}}            
\def\memras{\reff@jnl{MmRAS}}           
\def\mnras{\reff@jnl{MNRAS}}            
\def\pra{\reff@jnl{Phys.Rev.A}}         
\def\prb{\reff@jnl{Phys.Rev.B}}         
\def\prc{\reff@jnl{Phys.Rev.C}}         
\def\prd{\reff@jnl{Phys.Rev.D}}         
\def\prl{\reff@jnl{Phys.Rev.Lett}}      
\def\pasp{\reff@jnl{PASP}}              
\def\pasj{\reff@jnl{PASJ}}              
\def\qjras{\reff@jnl{QJRAS}}            
\def\skytel{\reff@jnl{S\&T}}            
\def\solphys{\reff@jnl{Solar~Phys.}}    
\def\sovast{\reff@jnl{Soviet~Ast.}}     
 \def\ssr{\reff@jnl{Space~Sci.Rev.}}     
\def\zap{\reff@jnl{ZAp}}                        
\def\nat{\reff@jnl{Nature}}             

\def\LaTeX{L\kern-.36em\raise.3ex\hbox{a}\kern-.15em
    T\kern-.1667em\lower.7ex\hbox{E}\kern-.125emX}

\def\deg{^\circ}

\begin{document}

\title[Spectra of Galactic components observed by \emph{WMAP}]{A determination of the Spectra of Galactic components observed by \emph{WMAP}}

\author[R.D. Davies et al.]{R.D. Davies,$^{1}$\thanks{E-mail: rdd@jb.man.ac.uk} C. Dickinson,$^{1,2}$\thanks{E-mail: cdickins@astro.caltech.edu} A.J. Banday,$^{3}$\thanks{E-mail:banday@MPA-Garching.MPG.DE} T.R. Jaffe,$^{3}$\thanks{E-mail:tjaffe@mpa-garching.mpg.de} K.M. G\'{o}rski,$^{2,4,5}$ R.J. Davis$^{1}$ \\
$^1$Jodrell Bank Observatory, University of Manchester, Macclesfield, Cheshire, SK11 9DL. \\
$^2$California Institute of Technology, Department of Astronomy, M/S 105-24, Pasadena, CA 91125, USA. \\
$^3$Max-Planck-Institut f\"{u}r Astrophysik, Karl-Schwarzschildstrasse 2, 85741 Garching bei M\"{u}nchen, Germany. \\
$^4$Jet Propulsion Laboratory, M/S 169-327, 4800 Oak Grove Drive, Pasadena, CA 91109, USA. \\
$^5$Warsaw University Observatory, Aleje Ujazdowskie 4, 00-478 Warszawa, Poland.}


\date{Received **insert**; Accepted **insert**}
       
\pagerange{\pageref{firstpage}--\pageref{lastpage}} 
\pubyear{}

\maketitle
\label{firstpage}


\begin{abstract}
\emph{WMAP} data when combined with ancillary data on free-free,
synchrotron and dust allow an improved understanding of the spectrum
of emission from each of these components.  Here we examine the sky
variation at intermediate latitudes using a cross-correlation technique. In particular, we compare the observed
emission in 15 selected sky regions to three ``standard'' templates.

The free-free emission of the diffuse ionised gas is fitted by a
well-known spectrum at K and Ka band, but the derived emissivity
corresponds to a  mean electron temperature of $\sim
4000-5000$~K. This is inconsistent with estimates from galactic H{\sc
ii} regions although a variation in the derived ratio of \ha\  to
free-free intensity by a factor of $\sim$2 is also found from region
to region. The origin of the discrepancy is unclear.

The anomalous emission associated with dust is clearly detected in
most of the 15 fields studied.  Fields that are only weakly
contaminated by synchrotron, free-free and CMB are studied;  the
anomalous emission correlates well with the Finkbeiner et al. (1999)
model 8 predictions (FDS8) at 94 GHz, with an effective spectral index
between 20 and 60~GHz, of $\beta \sim -2.85$.  Furthermore, the
emissivity varies by a factor of $\sim 2$ from cloud to cloud. A
modestly improved fit to the anomalous dust at K-band is provided by
modulating the template by an estimate of the dust colour temperature,
specifically FDS8$\times {\rm T}^n$.  We find a preferred value $n\sim
1.6$, although  there is a scatter from region to region.
Nevertheless, the preferred index drops to zero at higher frequencies
where the thermal dust emission dominates.

The synchrotron emission steepens between GHz frequencies and the
\emph{WMAP} bands. There are indications of spectral index variations
across the sky but the current data are not precise enough to
accurately quantify this from region-to-region.

Our analysis of the WMAP data indicates strongly that the
dust-correlated emission at the low WMAP frequencies has a spectrum
which is compatible with spinning dust;  we find no evidence for a
synchrotron component correlated with dust. The importance of these
results for the correction of CMB data for Galactic foreground
emission is discussed.
\end{abstract}



\setcounter{figure}{0} \normalsize

\section{INTRODUCTION}
\label{sec:introduction}

The all-sky observations by the Wilkinson Microwave Anisotropy Probe
(\emph{WMAP}: Bennett et al. 2003a) provide unprecedented data on
Galactic emission components in the frequency range 23 to 94 GHz, with
a high precision estimate of the CMB power spectrum. As CMB studies
move to higher precision it becomes necessary to determine the various
components of Galactic foreground emission to higher and higher
accuracy.  As an example of this requirement, the question of the
glitch in the power spectrum at multipole $\ell = 40$ (Hinshaw et
al.~2003)\footnote{The glitch at $\ell=40$ is still
present in the new 3-year \emph{WMAP} data (Hinshaw et al.~2006).} is
debated and various sources have been proposed including a Galactic
origin.  In an analysis of structures in the \emph{WMAP} CMB map
derived after removing Galactic foregrounds, Hansen, Banday \&
G\'{o}rski (2004) find a range of asymmetric structures on scales of
tens of degrees.  References to the many analyses which have detected
asymmetries or non-Gaussian structures may be found in this paper.

The role of a Galactic component in the above scenarios is unclear,
but cannot be ruled out.  Of particular relevance to this discussion
is the fact that each of the foreground components has a spectral
index that varies from one line of sight to another so using a single
spectral index can lead to significant uncertainties in the
corrections required.

It is obvious that the foregrounds that can be studied with
\emph{WMAP} data are of interest in their own right.  In comparing the
maps at the 5 frequencies of \emph{WMAP} (23, 33, 41, 61 and 94 GHz)
with the free-free, synchrotron dust templates it is possible to
clarify important properties of the emission.  For the free-free one
can derive the electron temperature distribution in the brighter
regions of the Galaxy (near the Galactic plane and in the Gould Belt
system).  In the case of the synchrotron emission significant
information on the spectral index variations across the sky can be
established.  Even more important, data are available to help clarify
the FIR-correlated emission.

The greatest insight is likely to come from \emph{WMAP} for the
dust-correlated emission.  The present situation is far from clear.
The dust-correlated emission was observed in the \emph{COBE}-DMR data
(Kogut et al. 1996) but was originally thought to be free-free. Leitch
et al. (1997) suggested the excess emission could be hot ($10^{6}$~K)
free-free emission. Draine \& Lazarian (1998a,b)
moved attention to the dust itself as the emission source through
dipole emission from spinning grains, referred to as ``spinning
dust''. They also considered an enhancement to the thermal emissivity
produced by thermal fluctuations in the grain magnetisation (Draine \&
Lazarian, 1999), but this explanation is less favoured by the data.
The lower frequency Tenerife results show that it was incompatible
with free-free (Jones et al.~2001) while the Tenerife data at 10, 15
and 30 GHz (de Oliveira-Costa et al. 1999, 2000, 2002) provided strong
evidence for dust at intermediate Galactic latitudes emitting a
spectrum of the form expected by spinning dust. A reanalysis of the
intermediate and high Galactic latitude data taken by \emph{COBE} and
supplemented by 19~GHz observations (Banday et al.~2003) led to
similar conclusions.  Finkbeiner, Langston \& Minter (2004) used 8.35
and 14.35~GHz data in combination with \emph{WMAP} data and found a
similar spectrum for a different environment in the Galactic ridge
(effectively $|b| < 4\deg$) in the inner Galaxy ($l \sim 15\deg$ to
$45\deg$);  the effect was not so clear-cut in the central regions of
the Galaxy ($|l| < 7.\!\deg 5$). The first targetted
search was carried out by Finkbeiner et
al. (2002) where they found a rising spectrum over the $5-10$~GHz
range for 2 diffuse clouds, which was interpreted as tentative
evidence for spinning dust. Finkbeiner (2004) also considers
intermediate latitudes in the \emph{WMAP} data in which he finds an
anomalous component compatible with spinning dust or a hot gas
($10^{6}$~K) component, but inconsistent with a traditional free-free
spectral index. New results from the Cosmosomas experiment (Watson et
al.~2005) also detect strong anomalous emission from the Perseus
molecular cloud with a rising spectrum in the range $11-17$~GHz,
suggestive of spinning dust. In contrast the \emph{WMAP} team (Bennett
et al.~2003b) gave a radically different interpretation of the
dust-correlated emission, considering it to be synchrotron emission
supposedly from star-forming regions associated with the dust;  again
this analysis was at intermediate and high Galactic latitudes.

Our present approach is to identify regions away from the Galactic
plane which are expected to be dominant in one of the three foreground
components, free-free, synchrotron or dust and to derive the spectrum
for each component. Five regions covering angular scales of $3\deg$ to
$20\deg$ were chosen for each component, based on foreground template
maps, making 15 in all. Such a selection is intended
to minimise the potential cross-talk between the various physical
components.  Moreover, by considering regions which are largely
dominated by well-known objects selected at a given frequency, it is
likely that the spectral behaviour is uniform over the region in
question thus supporting the use of a template based comparison.  We are
also interested in evaluating spectral variations over the sky, and
intend that any region to region scatter should reflect this. Two
complementary analyses of each region were considered.

The classical T--T plot approach can provide a detailed look at the
distribution of the data. In order to minimise cross-talk with the CMB
background, each of the five bands must be corrected for this
component as described by Bennett et al. (2003b) employing an internal
linear combination (ILC) of the \emph{WMAP} sky maps. For the high-latitude sky
considered here, this corresponds to a single set of linear
coefficients for each of the 5 frequencies. The ILC map at high
latitudes is therefore simply given by 0.109K $-$0.684Ka $-$0.096Q
$+$1.921V $-$0.25W. Unfortunately, subtracting the ILC CMB map
changes the relative levels of foreground emissions at each frequency
depending on the spectral characteristics of a given component.  This
``aliasing'' effect (see appendix\ref{sec:aliasing}), together with
other potential cross-talk between the foreground components, renders
the method useful only for visualisation and qualitative analysis.
Instead, all quantitative results in this paper are derived using a
cross-correlation (C--C) method, similar to the approach taken by
Banday et al. (2003).  The C--C analysis does not rely on a given CMB
map. Instead, the CMB is taken into account internally by including a
CMB component into the covariance matrix (see
section~\ref{sec:cc_method}), and the  the various correlations are
solved for simultaneously.

Section~\ref{sec:templates} describes the foreground templates used in
this analysis while Section~\ref{sec:fields} gives the considerations
for selecting the 15 regions for investigation. The
cross-correlation analysis and results are presented in
Section~\ref{sec:cc}.  The spectrum of each component is discussed
further in Section~\ref{sec:discussion}. A comparison with the new
3-year \emph{WMAP} data is described in Section~\ref{sec:wmap3} and
overall conclusions given in Section~\ref{sec:conclusions}.


\section{Templates used in the analysis}
\label{sec:templates}

The present analysis of the \emph{WMAP} data seeks to quantify the Galactic
foreground components of free-free, synchrotron and dust emission
using appropriate templates of each component.  The approach is
similar to that of Banday et al. (2003) in their analysis of the
\emph{COBE}-DMR data but with the difference that the current work relates to
selected areas rather than the full sky with the Galactic plane
removed.  Our analysis is made on an angular scale of $1\deg$, the
smallest that is feasible with the templates available, namely, $1
\deg$ in the \ha~free-free map, $0.\!^{\circ}85$ in the 408~MHz
synchrotron map and $0.\!^{\circ}82$ in the K-band of \emph{WMAP}. The basic properties
of the main maps used in the current analysis are summarised in
Table~\ref{tab:datasets}.


\subsection{\emph{WMAP} data}
\label{sec:wmap}

We use the 1st-year \emph{WMAP} data (Bennett et al. 2003a) provided in the
HEALPix\footnote{http://www.eso.org/science/healpix/.} pixelisation
scheme, with a resolution parameter of $N_{side}=512$, which can be
obtained from the LAMBDA
website\footnote{http://lambda.gsfc.nasa.gov/.}. The data consist of 5
full-sky maps covering the frequency range 23~GHz (K-band) up to
94~GHz (W-band); see Table~\ref{tab:datasets}. For our lower
resolution analysis and to compare to the foreground templates, these
maps (as well as all templates ) are smoothed to a common resolution
of $1\degr$ and converted to $\mu K$ of antenna temperature.  The
smoothed maps are then downgraded to a HEALPix resolution of
$N_{side}=128$, with a total of 196608 pixels.

We convert from thermodynamic temperature to brightness temperature
units i.e. the Rayleigh-Jeans convention\footnote{The conversion
factor from thermodynamic units to brightness (antenna) units (the
``Planck correction'') is given by the derivative of the Planck
function: $x^2e^{x}/(e^{x}-1)^2$ where $x=h \nu / k_{b}T_{\rm
CMB}$.}. This corresponds to a correction of $1.4$ per cent at 23~GHz
increasing to $25$ per cent at 94~GHz. Bright point sources are masked
using the mask templates provided by the \emph{WMAP} team. They are based on
various catalogues covering a wide range of wavelength domains masking
almost 700 sources in total (see Bennett et al.~2003b for details);
pixels within $0.\!^{\circ} 6$ radius of a source are blanked. This
operation typically removes $\sim 10$~per cent of the pixels in each
region. Fainter sources not included in the mask are not expected to
make a significant change in the results presented here. Any bright
sources still remaining would be easily identified in the maps.

The effective centre frequency of each band depends on the continuum
spectrum of the foreground being considered.  We adopt the values
given by Jarosik et al. (2003) which apply to the CMB blackbody
spectrum.  Reference to Page et al. (2003) shows that the effective
frequencies for synchrotron and free-free respectively are 1.0 and 0.7
per cent lower while those of the thermal (vibrational) dust are 0.7
per cent higher.  Thus using the frequencies appropriate for the CMB
will not have a significant affect on our estimates of spectral index
for the various foregrounds.

\begin{table}
\caption{\small Properties of the main maps used in the analysis. References are [1]: Haslam et al. (1982); [2]: Bennett et al. (2003a); [3]: FDS; [4]: Dickinson et al. (2003).}
\begin{tabular}{>{\small}l>{\small}c>{\small}c>{\small}c}
\hline
{\bf Dataset} & Frequency/  & Beamwidth            &{Reference} \\
              &Wavelength   &{\tiny (FWHM \degr)}  & \\    \hline
Haslam   & 408~MHz    & 0.85   & [1]  \\
\emph{WMAP} K   & 22.8~GHz   & 0.82   & [2]  \\
\emph{WMAP} Ka  & 33.0~GHz   & 0.62   & [2]  \\
\emph{WMAP} Q   & 40.7~GHz   & 0.49   & [2]  \\
\emph{WMAP} V   & 60.8~GHz   & 0.33   & [2]  \\
\emph{WMAP} W   & 93.5~GHz   & 0.21   & [2]  \\
FDS8 dust       &94~GHz      & 0.10   & [3]  \\
H$\alpha$ &656.2~nm & $\sim 1$ & [4] \\
\hline
\end{tabular}
\label{tab:datasets}
\end{table}


\subsection{The \ha~free-free template}
\label{sec:ha_template}

The only effective free-free template at the intermediate and high
Galactic latitudes used in the present study comes from
\ha~emission. In this analysis we use the all-sky \ha~template
described in Dickinson et al. (2003, hereafter DDD) which is a
composite of WHAM Fabry-Perot survey of the northern sky (Haffner et
al.~2003) which gives a good separation of the geocoronal \ha~emission
and of the SHASSA filter survey of the southern sky (Gaustad et
al.~2001).  Baseline effects may be significant in the SHASSA data
where information is lost on scales $> 10\deg$ due to geocoronal
emission. To correct for the Galactic gradient with latitude, a
baseline correction was applied assuming a cosecant law for
declinations further south ($-30\deg$) where WHAM data are not
present. On the angular scales of the WHAM data ($1\deg$) the
sensitivity of both surveys are comparable at $\sim 0.1$ Rayleigh (R).
Recently Finkbeiner (2003, hereafter F03) has produced an all-sky
\ha~map by including data from the VTSS filter survey (Dennison,
Simonetti \& Topasna 1998). This map contains structure down to 6
arcmin in scale, but has effectively variable resolution  due to the
different resolutions of the WHAM and SHASSA surveys. The differences
between the Dickinson et al. and the Finkbeiner maps are determined to
be less than 15 per cent in $R$ over the common power spectrum range
($\ell = 2-200$). The largest discrepancies are apparent near the
``cross-over'' region of the datasets where baseline levels have been
determined in a different way. In these regions, baseline
uncertainties are typically $\sim 1~R$. For the majority of the sky,
the baseline levels are tied to the WHAM data which contains baseline
uncertainties of $\ltsim 0.1$~R (Haffner et al.~2003). The \ha\
solutions do not change appreciably when the Finkbeiner \ha\ map is
used showing the similarity between the 2 templates.


When using the \ha~map as a template for the free-free emission it is
necessary to correct for the foreground dust absorption.  Dickinson et
al.~(2003) used the 100~$\mu$m map given by Schlegel, Finkbeiner \&
Davis (1998, hereafter SFD98) to estimate an absorption correction in
magnitudes at the \ha~wavelength of $A({\rm H}\alpha) = (0.0462 \pm
0.0035)D^{T}f_{d}$ where $D^{T}$ is the SFD temperature-corrected
100$~\mu$m intensity in MJy sr$^{-1}$ and $f_{d}$ is the fraction of
dust in front of the \ha~in the line of sight.   A value of $f_{d}
\sim 0.5$ expected under the assumption that the  ionised gas and dust
are coextensive along the line of sight (i.e. uniformly
mixed). Dickinson et al.~(2003) find $f_{d} \sim 0.3$ and accordingly
$A({\rm H}\alpha)$ is $<0.2$~mag over most of the intermediate and
high latitude sky where $D^{T} < 5$~MJy~sr$^{-1}$ ;  at latitudes
below $\sim 5\deg$ the absorption is too high to make a reasonable
estimate of the true \ha~intensity. Banday et al. (2003) use
\emph{COBE}-DMR and 19-GHz data to place a upper limit of $f_{d}
\ltsim 0.35$ assuming $T_{e}=7000$~K. This confirms that zero
correction is required for high Galactic latitudes ($|b| \gtsim
20^{\circ}$). It is worth noting that for the \emph{WMAP} 1-year
analysis (Bennett et al.~2003b), which uses the Finkbeiner \ha~map,
$f_{d}=0.5$ was adopted for the entire sky. At high Galactic
latitudes, the dust column density is small enough for this to have
almost negligible effect ($A({\rm H}\alpha) \ltsim 0.1$); the variance
of the \ha~map corrected by this value  is 20--30 per cent larger than
for an uncorrected map, depending on the galactic mask employed.
We therefore adopt the uncorrected template in the following analysis.

The conversion of dust-corrected \ha~intensities to emission measure
({\it EM} in units of cm$^{-6}$~pc) and then to free-free emission is
well-understood. The brightness temperature $T_{b}$ can be related to
EM using $T_{b} \propto T_{e}^{-0.35} \nu^{-2.1} \times EM$. It
requires a knowledge of the electron temperature $T_{e}$ of ionised
gas which varies as $\nu^{\sim0.7}$ in the conversion of \ha\
intensity to brightness temperature at microwave frequencies.  For the
\emph{WMAP} bands K, Ka, Q, V and W this corresponds to 11.4, 5.2,
3.3, 1.4 and 0.6~$\mu$K~R$^{-1}$ respectively at $T_{e}=8000$~K; see
Dickinson et al.~(2003) for details.

A number of estimates are available for $T_{e}$ in regions of the
Galaxy relevant to the present intermediate and high latitude study,
namely at galactocentric distances $R \sim R_{0}$.  Shaver et
al. (1983) used RRLs from Galactic H{\sc ii} regions to establish a
clear correlation of $T_{e}$ with $R$;  their result was

\begin{equation}
T_{e}~(R) = (3150 \pm 110) + (433 \pm 40)~R
\end{equation}

The following similar relationship was found by Paladini, Davies \& De~Zotti~(2004)
from a larger sample which contained many weaker sources

\begin{equation}
T_{e}~(R) = (4170 \pm 120) + (314 \pm 20)~R
\end{equation}

At $R \sim R_{0}$, in the local region, these expressions indicate
that $T_{e} = (7200 \pm 1200)$~K.  It is possible that the $T_{e}$ of
diffuse H{\sc ii} emission at a given galactocentric distance may be
different from that of the higher density H{\sc ii} regions on the
Galactic plane.  There are strong indications from observation and
theory that the diffuse ionised gas will have a higher electron
temperature than in the density bounded H{\sc ii} regions which
contain the ionising stars (Wood \& Mathis 2004). RRLs give another
route to identifying the free-free component of the Galactic
foreground and may be useful at low Galactic latitudes when the
\ha\ signal is heavily absorbed by foreground dust.


\begin{figure*}
\begin{center}
\includegraphics[width=0.95\textwidth,angle=0]{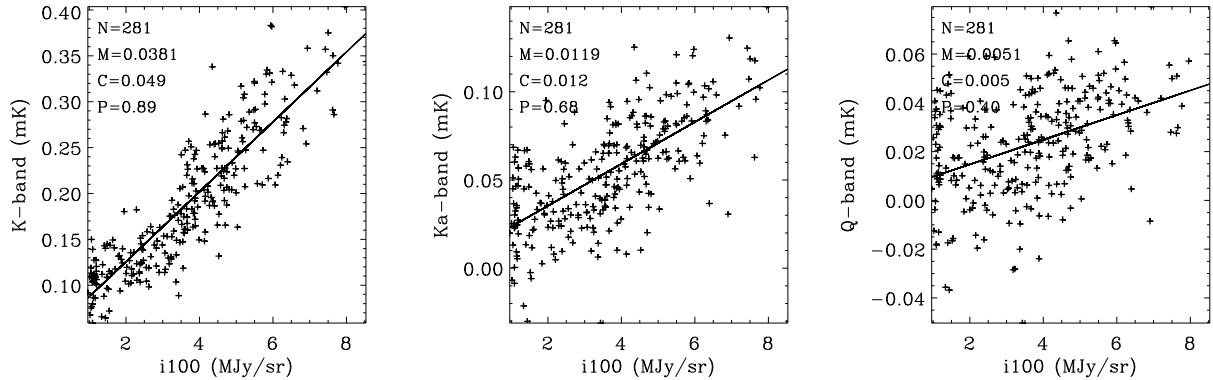}
 \caption{T--T plots for region 6 (middle row of
 Fig.~\ref{fig:region_example_maps}) clearly showing the \emph{WMAP}
 dust-correlated emission at K, Ka and Q bands against the $100~\mu$m
 SFD98 map. The best-fitting line is plotted along with the number of
 pixels N, the y-intercept C, slope M and Pearson correlation
 coefficient P. \label{fig:tt_plots_region6}}
\end{center}
\end{figure*}

\subsection{The dust template}

Dust has two broadband emission components in the frequency range 10
to 1000~GHz.  The anomalous emission is dominant at the lower end
while the thermal (vibrational) component is responsible for the
higher end. We will see that the anomalous component is the strongest
for the \emph{WMAP} frequencies of 23, 33 and 41 GHz while the
vibrational component dominates at 94 GHz.  We clearly need templates
for both components if we are to quantify accurately the anomalous
emission at 61 and 94 GHz.

\emph{COBE}-DIRBE full-sky maps at 100, 140 or 240~$\mu$m  with
$0.\!^{\circ}7$ resolution have commonly been used as tracers of the
thermal dust component (Kogut et al. 1996).  However, the most
sensitive full-sky map of dust emission is the 100~$\mu$m data at
6~arcmin resolution from $IRAS$.  These data have been recalibrated
using \emph{COBE}-DIRBE data and reanalyzed to give reduced artifacts
due to zodiacal emission and to remove discrete sources (SFD98). In a
preliminary analysis, we utilised the latter to help define specific
dust fields of interest, and then to examine the dust emissivity of
the 15 selected  regions (see section~\ref{sec:fields}).
Fig.~\ref{fig:tt_plots_region6} shows T--T plots for one of the
regions in K, Ka and Q-bands of \emph{WMAP} against the $100~\mu$m
map. The dust-correlation is striking, particularly at K- and
Ka-bands.  Moreover, the emissivity ($\mu$K/(MJy~sr$^{-1}$)) was
determined  to vary by a factor of $\sim 2.5$ from cloud to cloud.
However, this scatter was found, at least in part, to
be driven by variations in the dust temperature. Finkbeiner, Davis \&
Schlegel (1999; hereafter FDS) recognised the importance of this for
predictions of the dust contribution at microwave wavelengths, and
developed a series of models based on the 100 and 240 $\mu$m  maps
tied to \emph{COBE}-FIRAS spectral data in the range 0.14 to
3.0~mm. The preferred model 8 (hereafter FDS8) has a spectral index
$\beta \approx +1.7$ over the \emph{WMAP} frequencies. For the work
undertaken in this paper, we adopt the FDS8 predicted emission at 94
GHz as our reference  template for dust emission. Note
that in previous work, correlations were often referenced to the SFD98
$100~\mu$m template, in units of $\mu$K/(MJy~sr$^{-1}$). To convert
these values to correlations relative to FDS8, they should be divided
by $\sim 3.3$.


\subsection{The synchrotron template}

The synchrotron emission of the Galaxy is best studied at low
frequencies ($<1$~GHz) where it is least contaminated by other
emission (principally free-free emission from the ISM).  Studies at
these frequencies show that the temperature spectral index ($T_{b}
\propto \nu^{\beta}$) has typical values of $\beta = -2.55$ and $
-2.8$~(Lawson et al.~1987) at 38 and 800 MHz respectively.  Reich \&
Reich (1988) demonstrated a range of spectral index values
$\beta=2.3-3.0$ between 408 and 1420 MHz, with a typical dispersion
$\Delta \beta= \pm 0.15$.  At higher frequencies $\beta$ is expected
to increase by $\sim 0.5$ due to radiation losses in the relativistic
CR electrons responsible for the synchrotron emission.

We use the 408 MHz map by Haslam et al.~(1981, 1982) which is the only
all-sky map with adequate resolution (51 arcmin) at a sufficiently low
frequency.  It has a brightness temperature scale which is calibrated
with the 404 MHz $8.\!^{\circ}5 \times 6.\!^{\circ}5$ survey of
Pauliny-Toth and Shakeshaft (1962).  
The 1.4~GHz northern sky map with
a resolution of 35 arcmin made by Reich \& Reich (1986) and the
2.3~GHz map at a resolution of 20 arcmin from Jonas, Baart \&
Nicholson (1998) are employed to provide frequency coverage at GHz
frequencies when assessing the spectral index of emission regions
selected in the present study.  
We note that spurious baseline effects
have been identified in these surveys (Davies, Watson \& Guti\'{e}rrez
(1996)) which can affect $\beta$ determinations for weaker features.
In the current study we select stronger emission regions for
comparison with the \emph{WMAP} data;  such strong regions are essential when
extending the spectra to the highest map frequencies (94~GHz) where
$\beta \sim -3.0$.


\begin{table*}
\centering
\caption{Summary of 15 selected regions of sky.  \label{tab:field_list}}
\begin{tabular}{ccccl}
\hline
Field   &Dominant     &Longitude           &Latitude             &Description.  \\
No.     &Emission     &Range               &Range                &       \\ \hline

1       &Free-free    &$245\deg - 260\deg$ &$+21\deg - +31\deg$
&Northern edge of Gum Nebula.       \\ 2       &Free-free    &$140\deg
- 155\deg$ &$+15\deg - +20\deg$  &Disc-like structure above Galactic
plane.       \\ 3       &Free-free    &$200\deg - 230\deg$ &$-41\deg -
-48\deg$  &Eridanus complex - within southern Gould Belt system.
\\ 4       &Free-free    &$250\deg - 260\deg$ &$-25\deg - -35\deg$
&Southern edge of Gum Nebula       \\ 5       &Free-free    &$90\deg -
97\deg$   &$-13\deg - -30\deg$  &Below plane in northern sky.       \\
\hline 6       &Dust         &$118\deg - 135\deg$ &$+20\deg - +37\deg$
&$l=125\deg$ dust spur, NCP region (the ``duck'').       \\ 7       &Dust         &$300\deg -
315\deg$ &$+35\deg - +45\deg$  &Outer edge of northern Gould Belt
system.       \\ 8       &Dust         &$227\deg - 237\deg$ &$+12\deg
- +18\deg$  &Above plane in southern sky.       \\ 9       &Dust
&$145\deg - 165\deg$ &$-30\deg - -38\deg$  &Orion region in southern
Gould Belt.       \\ 10      &Dust         &$300\deg - 320\deg$
&$-30\deg - -40\deg$  &Below plane southern sky.       \\ \hline 11
&Synchrotron  &$33\deg - 45\deg$   &$+50\deg - +70\deg$  &Middle
section of North Polar Spur.       \\ 12      &Synchrotron
&$270\deg - 310\deg$ &$+55\deg - +70\deg$  &Outermost section of North
Polar Spur.       \\ 13      &Synchrotron  &$350\deg - 5\deg$
&$-35\deg - -50\deg$  &Southern bulge in synchrotron sky.       \\ 14
&Synchrotron  &$70\deg - 90\deg$   &$+20\deg - +30\deg$  &A ``weak''
northern spur.       \\ 15      &Synchrotron  &$76\deg - 84\deg$
&$-30\deg - -50\deg$  &A southern spur.       \\ \hline
\end{tabular}
\end{table*}

\section{Field selection}
\label{sec:fields}

\begin{figure}
\begin{center}
\includegraphics[width=0.5\textwidth,angle=0]{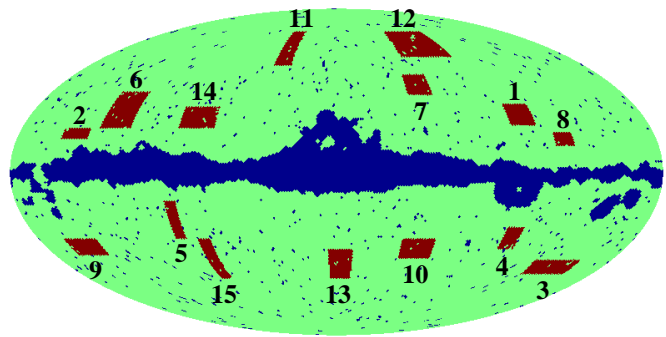}
  \caption{Mollweide full-sky map showing the 15 selected regions in
  red with the masked regions in blue, based on the Kp2 Galactic plane
  mask and $\sim 700$ bright radio sources. The Galactic centre is in
  the middle and longitude increases to the left. \label{fig:15_field_locations}}
\end{center}
\end{figure}

The fields selected for study were chosen on the basis that one of the
3 foregrounds (free-free, dust or synchrotron emission) was dominant
in each field.  This criterion can be satisfied at intermediate
Galactic latitudes, well away from the Galactic plane where the
foregrounds are inevitably confused.  As a result, our study will
sample conditions in the Local Arm or adjacent spiral arms.

The angular scale of this study was determined by the largest
beamwidths of critical elements in the data sets.  The beamwidth of
the 408~MHz survey is 51 arcmin, that of the WHAM \ha~survey is
$1\deg$ and that of the \emph{WMAP} lowest frequency, 22.8~GHz, is
49~arcmin (Table~\ref{tab:datasets}). All the other data sets had a
higher resolution.  Accordingly a beamwidth of $1.\!\deg 0$ was chosen
as the appropriate resolution and the analysis of the data sets was
undertaken by smoothing to this resolution.

The choice of a $1.\!\deg 0$ resolution determined the size of
structures in the \emph{WMAP} maps which could be studied to best
effect.  The best signal-to-noise ratio would be achieved in
structures on a scale of several resolution elements which therefore
contain a number of independent data points.  Many of the features
would be elongated or contain structure.  Another selection
requirement was that the field should contain a smooth background
covering approximately half the area.  This was essential in
identifying the feature and separating its emission from underlying
emission.  The features studied typically had structure on scales of
$3\deg$ to $10\deg$.

5 fields were chosen in which each of the three Galactic foregrounds
were dominant.  The synchrotron fields were selected from the Haslam
et al. (1982) 408~MHz map, the free-free fields from the Dickinson et
al. (2003) \ha\ map and the dust fields from the SFD98 $100~\mu$m
map. Table~\ref{tab:field_list} lists the 15 fields of the present
study.  For each field the dominant emission and the Galactic
coordinates are given along with a short description of the field.

\begin{figure*}
\begin{center}
\includegraphics[width=1.0\textwidth,angle=0]{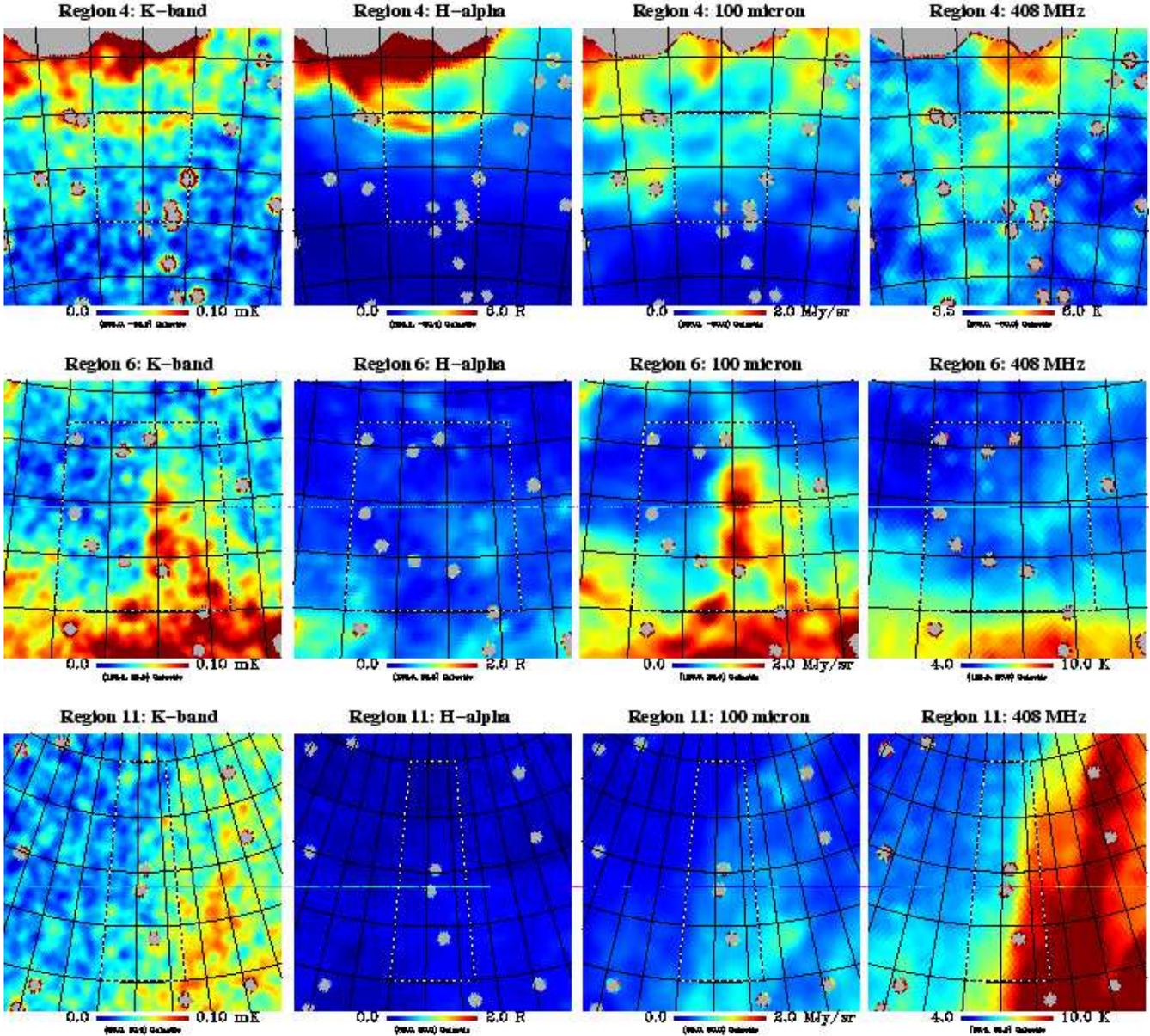}
 \caption{Maps of region 4 (\ha\ (free-free) dominated; {\it top}
 row), region 6 (dust dominated; {\it middle} row), and region 11
 (synchrotron dominated; {\it bottom} row). From left to right are
 maps at \emph{WMAP} K-band, \ha, SFD98 $100~\mu$m dust intensity and
 408~MHz. Galactic coordinates are shown. Each map, with a pixel
 resolution $N_{side}=256$, covers a $25\deg \times 25\deg$ area with
 $1\deg$ resolution. The dotted black/white line delineates the actual
 areas used for the T--T plots and cross-correlation
 analyses. Grey-areas are the standard \emph{WMAP} Kp0 mask and
 extragalactic sources mask. \label{fig:region_example_maps}}
\end{center}
\end{figure*}

Fig.~\ref{fig:15_field_locations} shows the position of the 15
selected  regions overlaid on the Kp2 intensity mask and source mask
($\sim 700$ sources in total) used by the \emph{WMAP} team (Bennett et
al.~2003b). Fig.~\ref{fig:region_example_maps} shows 3 regions
(regions 4,6 and 11) with an ILC-subtracted K-band, \ha, $100~\mu$m
and 408~MHz data. The dominant foreground in each region (see
Table~\ref{tab:field_list}) is clearly seen along with the correlated
emission at K-band.

\section{Cross-correlation analysis}
\label{sec:cc}

The cross--correlation (C--C) method used here is a least--squares fit
of one map to one or more templates. In the absence of any covariance
information on the residuals, and assuming no net offsets (\ie
monopoles), this method is equivalent to the classical T--T method
when only one template map ($T_1\equiv{\mathbf t}$) is compared to the
data ($T_2\equiv{\mathbf d}$). The advantage of the C--C method is
that we can fit several components simultaneously and that we can
include information about the CMB  thorough its signal covariance
rather than having to  correct for it. The issues of
CMB subtraction and correlated components are discussed further in
Appendices A \& B.


\subsection{Method}
\label{sec:cc_method}

The cross--correlation measure, $\alpha$, between a data vector,
${\bf d}$ and a template vector ${\bf t}$ can be measured by
minimising:

\begin{equation}
\chi^2 = ({\bf d}-\alpha {\bf t})^T \cdot {\bf M}^{-1}_{SN}\cdot ({\bf d}-\alpha {\bf t}) = {\bf \tilde{d}}^T \cdot {\bf M}^{-1}_{SN}\cdot {\bf \tilde{d}}
\end{equation}

where ${\bf M}_{SN}$ is the covariance matrix including both signal
and noise for the template--corrected data vector ${\bf \tilde{d}}
\equiv {\bf d} - \alpha {\bf t}$.  Solving for $\alpha$ then becomes:

\begin{equation}
\alpha = \frac{ {\bf t}^T\cdot{\bf M}^{-1}_{SN}\cdot {\bf d} }{ {\bf t}^T\cdot{\bf M}^{-1}_{SN}\cdot {\bf t} }
\label{eq:cc_basic}
\end{equation}

To compare multiple template components ${\mathbf t}_j$, {\it e.g.},
different foregrounds, to a given dataset, the problem becomes a
matrix equation. G\'{o}rski et al. (1996) describe the method in
harmonic space, which is fundamentally no different from pixel space.
In the case where we have $N$ different foreground components, we end
up with the simple system of linear equations ${\bf Ax}={\bf b}$,where
\begin{gather}
A_{kj}=\mathbf{t}^T_k \cdot \mathbf{M}_{\textrm{SN}}^{-1} \cdot
\mathbf{t}_j, \notag \\ b_k = \mathbf{t}^T_k \cdot
\mathbf{M}_{\textrm{SN}}^{-1} \cdot \mathbf{d} \notag, \\ x_k
=\alpha_k.
\end{gather}
When only one template is present, this reduces to equation
(\ref{eq:cc_basic}) above.

The signal covariance is that for theoretical CMB anisotropies,
$M^S_{ij} = \frac{1}{4\pi} \sum^\infty_{\ell=0} (2\ell+1) C_{\ell}
B^2_{\ell} P_{\ell}(\hat{n}_i\cdot\hat{n}_j) $, where $B_{\ell}$ is
the Gaussian beam of $1\degr$ FWHM.  The power spectrum, $C_{\ell}$,
is taken from the \emph{WMAP} best fit $\Lambda$CDM power law spectrum
(Bennett et al.~2003a).  The noise covariance is determined from the
uncorrelated pixel noise as specified for each pixel in the
\emph{WMAP} data, and subsequently convolved as described above.

For each of the fifteen regions, the data vector includes only those
pixels of interest, and the covariance matrices are only the
corresponding rows and columns.  These regions vary in size from
230 to 1140 pixels at this resolution.

The errors $\delta\alpha_{k}$ are the square root of the diagonal of
${\bf A}^{-1}$. The simultaneous fitting of multiple template
components allows us to deal with the fact that, though the regions
are chosen to be dominant in one given component, they are not
entirely free of the other components.
The simultaneous fitting of multiple foreground components allows such
cross--talk to be quantified.


\subsection{Results of the cross-correlation analysis}
\label{sec:cc_results}

At each \emph{WMAP} band the emissivity of the 3 foreground components
(free-free, dust and synchrotron) has been estimated as a ratio of
template brightness; 
$\mu$K~R$^{-1}$, $\mu$K/$\mu$K$_{\rm FDS8}$ and $\mu$K~K$^{-1}$
respectively.
The analysis was a joint solution derived for all 3 components
simultaneously. For each of the components we also made solutions for
the all-sky (Kp2 cut) \emph{WMAP} data.


\subsection{Free-free emission}
\label{sec:free-free}

The 5 isolated regions with strong \ha~features ($1-5$) are listed in
Table~\ref{tab:field_list}. Free-free emission should in principle be
detectable at radio frequencies from 408~MHz up to and including the
\emph{WMAP} frequencies. Accordingly we have used the additional radio
surveys at 1.4~GHz (Reich \& Reich, 1988) and 2.3~GHz (Jonas et
al. 1998) to confirm that the regions exhibit a general flattening of
spectral index between 408~MHz and 2.3~GHz suggesting that there  is
considerable free-free emission, at least relative to any synchrotron
component at these frequencies.  The peak \ha\ intensities in the 5
maps are in the range $10 - 40~R$. The results of the analysis for K-
and Ka-bands, where free-free emission will be strongest, are given in
Table~\ref{tab:ff_ratios}. For comparison we list the fits for the two
\ha~templates (DDD and F03) and find overall, there is good agreement
between them. The results indicate a lower electron temperature of
roughly $T_{e}=4000-5000$~K rather than the often-assumed
$T_{e}=8000$~K.  However, it is important to note that there is
variation in this ratio by a factor of $\sim$2 from region to
region. The average of the 5 fields is consistent with the high
Galactic latitude solution (Kp2 cut).

%
\begin{table}
\begin{tabular}{ccccc}
\hline
Field & Template & $\frac{T_{\rm K}}{T_{\rm H_\alpha}}$  & $\frac{T_{\rm Ka}}{T_{\rm H_\alpha}}$ & H$_\alpha$ intensity \\
No. &  & $\mu$K R$^{-1}$ & $\mu$K R$^{-1}$ & range R  \\
\hline
\multirow{2}{.02\hsize}{1} 
 &  F03 &  11.3$^{\pm  4.8}$ &   9.8$^{\pm  4.4}$ & 1-10\\
 & DDD &   9.8$^{\pm  4.4}$ &   6.2$^{\pm  4.1}$\\
\multirow{2}{.02\hsize}{2} 
 &  F03 &   4.7$^{\pm  3.0}$ &   0.7$^{\pm  2.8}$ & 3-15\\
 & DDD &   5.2$^{\pm  2.9}$ &   1.4$^{\pm  2.7}$\\
\multirow{2}{.02\hsize}{3} 
 &  F03 &   5.5$^{\pm  3.0}$ &   3.1$^{\pm  2.6}$ & 2-14\\
 & DDD &   2.3$^{\pm  3.5}$ &  -1.7$^{\pm  3.2}$\\
\multirow{2}{.02\hsize}{4} 
 &  F03 &   7.4$^{\pm  2.6}$ &   4.4$^{\pm  2.3}$ & 3-16\\
 & DDD &   7.2$^{\pm  2.1}$ &   1.2$^{\pm  1.1}$\\
\multirow{2}{.02\hsize}{5} 
 &  F03 &   9.6$^{\pm  1.1}$ &   4.8$^{\pm  1.0}$ & 3-40\\
 & DDD &  10.1$^{\pm  1.2}$ &   5.1$^{\pm  1.1}$\\
\hline
\multirow{2}{.02\hsize}{\small Avg.} 
 & F03 &   8.6$^{\pm  0.9}$ &   4.4$^{\pm  0.8}$ \\
 & DDD &   8.5$^{\pm  0.9}$ &   3.0$^{\pm  0.7}$ \\
\hline
\multirow{2}{.02\hsize}{\small Kp2}
 & F03 &   7.7$^{\pm  0.9}$ &   3.7$^{\pm  0.9}$ \\
 &  DDD &   7.5$^{\pm  0.9}$ &   3.6$^{\pm  0.9}$ \\
\hline
 & T$_e$=4000K &   8.0 &   3.6\\
 & T$_e$=5000K &   8.9 &   4.1\\
 & T$_e$=6000K &   9.8 &   4.5\\
 & T$_e$=7000K &  10.6 &   4.9\\
 & T$_e$=8000K &  11.4 &   5.2\\
\hline

\end{tabular}
\caption{ Free-free emission as determined by C--C method for K-and Ka-band \emph{WMAP} data.  Units are brightness temperature T ($\mu$K) relative to unit H$\alpha$ intensity $I_{H_\alpha}$ (R).  The expected values are shown at the bottom from the predictions given by Dickinson et al. (2003) assuming different electron temperatures.  The range of H$\alpha$ intensities in each region is also given in the last column.  For each region, two templates are used: the Finkbeiner (2003) template, F03, or the Dickinson et al. (2003), DDD.  Full sky fits, outside the Kp2 cut, are also shown; here, only a diagonal approximation to the covariance matrix is used, and the uncertainties are determined using simulations (see Appendix~\ref{appendix_cc}) and the DDD template only. \label{tab:ff_ratios} }
\end{table}


\subsection{Anomalous dust emission}
\label{sec:anom_dust}

The dust emissivity for all fields for the 5 \emph{WMAP} frequency
bands is given in Table~\ref{tab:anomalous_dust}. The emissivity is
that relative to the FDS8 prediction for the W-band and is given
separately for the raw \emph{WMAP} data and for \ha\ corrected
data. Also shown is the range of dust temperature in each field taken
from SFD98.  The five fields ($6-10$) that were selected on the basis
that they exhibited  dust emission only weakly confused by synchrotron
or \ha\ (free-free) emission produced the most significant
correlations amongst the 15 fields studied.  In addition field numbers
2,5,11,14 and 15 also show significant dust correlations; this is
because the anomalous dust emission is the dominant foreground at the
lower \emph{WMAP} frequencies ($20-40$~GHz). However, in these latter
fields some confusion from synchrotron and free-free might have been
expected. The possible effect of a contribution from free-free
emission has been tested by subtracting the \ha\ template converted to
free-free brightness temperature with an electron temperature $T_{e}=
4000$~K, consistent with our previous findings. The free-free
correction is small relative to the dust and the dust results remain
remarkably robust.


It can be seen from Table~\ref{tab:anomalous_dust} that there is a
spread of a factor of $\approx 2$ in the emissivity of dust
clouds. The range of dust emission spectral index as determined for
example by the ratio of K to Ka-band emissivity of individual clouds
is less than this ($\sim 1.4$ in the dust-dominated regions) and may
also be a significant result.

The spectrum of the the anomalous dust emissivity is best given by the
average of the clouds listed in Table~\ref{tab:anomalous_dust}. The
average values are seen to be slightly higher than that of the
full-sky (Kp2 cut). Note that the traditional vibrational dust
component is negligible in the K, Ka and Q-bands and is only dominant
at W-band.

The anomalous dust emissivity, when corrected for vibrational
emission, shows an average spectral index $\beta$ of $-2.85$ (varying
from $\approx -2.8$ to $-3.8$ in the dust-dominated regions) and is
discussed further in section~\ref{sec:dust_discussion}.

%
\begin{table*}
\begin{tabular}{ccccccccc}
\hline
\small Field & FDS range & Dust T range & \multicolumn{5}{c}{\small Dust emissivity relative to FDS } & Notes \\
\small No. & ($\mu$K) & (K) & K & Ka & Q  & V & W & \\
\hline
\multirow{2}{0.02\hsize}{1} & \multirow{2}{0.08\hsize}{ 4.4-10.3} & \multirow{2}{0.08\hsize}{17.6-18.5}   &   8.4$^{\pm  5.9}$ &  -0.1$^{\pm  5.1}$ &   2.4$^{\pm  4.3}$ &  -1.0$^{\pm  3.3}$ &  -2.3$^{\pm  2.6}$\\
 & &  &   8.9$^{\pm  5.7}$ &   0.5$^{\pm  5.0}$ &   3.4$^{\pm  4.2}$ &  -0.3$^{\pm  3.2}$ &  -1.6$^{\pm  2.5}$\\
\multirow{2}{0.02\hsize}{2} & \multirow{2}{0.08\hsize}{ 14.2-41.3} & \multirow{2}{0.08\hsize}{16.5-17.5}   &   6.6$^{\pm  1.6}$ &   1.8$^{\pm  1.3}$ &   2.0$^{\pm  1.2}$ &  -0.8$^{\pm  1.1}$ &   0.3$^{\pm  0.9}$\\
 & &  &   6.2$^{\pm  1.6}$ &   0.9$^{\pm  1.3}$ &   0.5$^{\pm  1.2}$ &  -2.1$^{\pm  1.0}$ &  -0.3$^{\pm  0.9}$\\
\multirow{2}{0.02\hsize}{3} & \multirow{2}{0.08\hsize}{ 1.6-10.1} & \multirow{2}{0.08\hsize}{17.6-18.5}   &  12.1$^{\pm  5.2}$ &   5.5$^{\pm  4.5}$ &   7.2$^{\pm  3.9}$ &   4.8$^{\pm  3.5}$ &  -0.7$^{\pm  3.1}$\\
 & &  &   7.7$^{\pm  4.7}$ &   1.6$^{\pm  4.1}$ &   3.8$^{\pm  3.6}$ &   2.7$^{\pm  3.3}$ &  -0.8$^{\pm  2.9}$\\
\multirow{2}{0.02\hsize}{4} & \multirow{2}{0.08\hsize}{ 2.5-10.4} & \multirow{2}{0.08\hsize}{17.8-18.3}   &   3.9$^{\pm  6.1}$ &   7.3$^{\pm  5.1}$ &   6.4$^{\pm  4.3}$ &   2.5$^{\pm  3.6}$ &  -2.0$^{\pm  3.0}$\\
 & &  &   2.9$^{\pm  5.6}$ &   4.8$^{\pm  4.9}$ &   5.0$^{\pm  4.3}$ &   2.0$^{\pm  3.5}$ &  -2.3$^{\pm  3.0}$\\
\multirow{2}{0.02\hsize}{5} & \multirow{2}{0.08\hsize}{ 6.9-30.2} & \multirow{2}{0.08\hsize}{17.2-18.6}   &  12.6$^{\pm  2.1}$ &   2.6$^{\pm  1.8}$ &   2.6$^{\pm  1.6}$ &  -0.3$^{\pm  1.4}$ &  -0.8$^{\pm  1.2}$\\
 & &  &  13.1$^{\pm  2.0}$ &   2.5$^{\pm  1.8}$ &   1.7$^{\pm  1.6}$ &  -1.6$^{\pm  1.4}$ &  -1.6$^{\pm  1.2}$\\
\hline
\multirow{2}{0.02\hsize}{\bf 6} & \multirow{2}{0.08\hsize}{\bf 2.8-51.1} & \multirow{2}{0.08\hsize}[-1.5mm]{\bf 15.7-18.1 }  & {\bf   6.7$^{\pm  0.7}$} & {\bf   2.9$^{\pm  0.6}$} & {\bf   2.1$^{\pm  0.6}$} & {\bf   0.8$^{\pm  0.5}$} & {\bf   2.3$^{\pm  0.4}$}\\
 & &  & {\bf   6.5$^{\pm  0.7}$} & {\bf   2.6$^{\pm  0.6}$} & {\bf   1.9$^{\pm  0.6}$} & {\bf   0.8$^{\pm  0.5}$} & {\bf   2.2$^{\pm  0.4}$}\\
\multirow{2}{0.02\hsize}{\bf 7} & \multirow{2}{0.08\hsize}{\bf 6.2-17.0} & \multirow{2}{0.08\hsize}[-1.5mm]{\bf 17.9-18.7 }  & {\bf   8.1$^{\pm  3.9}$} & {\bf   2.8$^{\pm  3.5}$} & {\bf   0.2$^{\pm  3.2}$} & {\bf   0.6$^{\pm  2.9}$} & {\bf   0.5$^{\pm  2.6}$}\\
 & &  & {\bf   7.8$^{\pm  3.8}$} & {\bf   2.8$^{\pm  3.5}$} & {\bf  -0.3$^{\pm  3.2}$} & {\bf   0.4$^{\pm  2.9}$} & {\bf   0.2$^{\pm  2.6}$}\\
\multirow{2}{0.02\hsize}{\bf 8} & \multirow{2}{0.08\hsize}{\bf 6.4-25.8} & \multirow{2}{0.08\hsize}[-1.5mm]{\bf 17.1-18.0 }  & {\bf   8.4$^{\pm  2.5}$} & {\bf   2.6$^{\pm  2.3}$} & {\bf   2.4$^{\pm  2.2}$} & {\bf   2.4$^{\pm  2.0}$} & {\bf   3.1$^{\pm  1.8}$}\\
 & &  & {\bf   8.2$^{\pm  2.5}$} & {\bf   2.6$^{\pm  2.3}$} & {\bf   2.5$^{\pm  2.2}$} & {\bf   2.5$^{\pm  2.0}$} & {\bf   3.1$^{\pm  1.8}$}\\
\multirow{2}{0.02\hsize}{\bf 9} & \multirow{2}{0.08\hsize}{\bf 9.6-97.5} & \multirow{2}{0.08\hsize}[-1.5mm]{\bf 15.4-17.8 }  & {\bf   7.3$^{\pm  0.6}$} & {\bf   2.9$^{\pm  0.6}$} & {\bf   1.8$^{\pm  0.5}$} & {\bf   1.4$^{\pm  0.5}$} & {\bf   1.9$^{\pm  0.4}$}\\
 & &  & {\bf   7.3$^{\pm  0.6}$} & {\bf   2.9$^{\pm  0.6}$} & {\bf   1.8$^{\pm  0.5}$} & {\bf   1.4$^{\pm  0.5}$} & {\bf   1.9$^{\pm  0.4}$}\\
\multirow{2}{0.02\hsize}{\bf 10} & \multirow{2}{0.08\hsize}{\bf 3.5-31.7} & \multirow{2}{0.08\hsize}[-1.5mm]{\bf 16.8-18.3 }  & {\bf  12.0$^{\pm  1.4}$} & {\bf   4.7$^{\pm  1.2}$} & {\bf   1.8$^{\pm  1.1}$} & {\bf   0.2$^{\pm  1.0}$} & {\bf   1.3$^{\pm  0.9}$}\\
 & &  & {\bf  12.2$^{\pm  1.3}$} & {\bf   5.0$^{\pm  1.2}$} & {\bf   1.8$^{\pm  1.1}$} & {\bf   0.3$^{\pm  1.0}$} & {\bf   1.3$^{\pm  0.9}$}\\
\hline
\multirow{2}{0.02\hsize}{11} & \multirow{2}{0.08\hsize}{ 1.9-7.6} & \multirow{2}{0.08\hsize}{17.6-18.3}   &  19.0$^{\pm  7.9}$ &   9.9$^{\pm  6.8}$ &   6.6$^{\pm  6.1}$ &   3.1$^{\pm  5.4}$ &   1.4$^{\pm  4.7}$\\
 & &  &  17.8$^{\pm  7.8}$ &   8.3$^{\pm  6.7}$ &   7.0$^{\pm  6.1}$ &   3.5$^{\pm  5.4}$ &   1.3$^{\pm  4.7}$\\
\multirow{2}{0.02\hsize}{12} & \multirow{2}{0.08\hsize}{ 2.2-5.9} & \multirow{2}{0.08\hsize}{17.6-18.2}   &  15.6$^{\pm  7.5}$ &   5.3$^{\pm  6.6}$ &   4.6$^{\pm  5.8}$ &  -7.1$^{\pm  5.0}$ &  -2.1$^{\pm  4.2}$\\
 & &  &  15.4$^{\pm  7.5}$ &   5.1$^{\pm  6.6}$ &   4.4$^{\pm  5.8}$ &  -7.3$^{\pm  5.0}$ &  -2.2$^{\pm  4.2}$\\
\multirow{2}{0.02\hsize}{13} & \multirow{2}{0.08\hsize}{ 1.9-10.5} & \multirow{2}{0.08\hsize}{17.5-18.5}   &   3.9$^{\pm  6.4}$ &  -3.7$^{\pm  5.7}$ &   1.6$^{\pm  4.9}$ &  -2.2$^{\pm  3.8}$ &  -1.0$^{\pm  3.1}$\\
 & &  &   3.6$^{\pm  6.3}$ &  -3.0$^{\pm  5.7}$ &   1.5$^{\pm  4.9}$ &  -2.2$^{\pm  3.8}$ &  -1.0$^{\pm  3.1}$\\
\multirow{2}{0.02\hsize}{14} & \multirow{2}{0.08\hsize}{ 3.7-9.8} & \multirow{2}{0.08\hsize}{17.5-18.5}   &   8.3$^{\pm  4.8}$ &   7.6$^{\pm  3.9}$ &   1.8$^{\pm  3.0}$ &   3.0$^{\pm  2.3}$ &  -0.6$^{\pm  1.8}$\\
 & &  &   7.1$^{\pm  4.7}$ &   7.1$^{\pm  3.9}$ &   1.1$^{\pm  3.0}$ &   2.2$^{\pm  2.3}$ &  -0.7$^{\pm  1.8}$\\
\multirow{2}{0.02\hsize}{15} & \multirow{2}{0.08\hsize}{ 6.1-16.3} & \multirow{2}{0.08\hsize}{17.2-18.3}   &   8.1$^{\pm  3.3}$ &  -1.0$^{\pm  3.0}$ &  -1.0$^{\pm  2.7}$ &  -1.3$^{\pm  2.5}$ &  -0.5$^{\pm  2.2}$\\
 & &  &   8.1$^{\pm  3.3}$ &  -0.8$^{\pm  3.0}$ &  -0.9$^{\pm  2.7}$ &  -1.4$^{\pm  2.5}$ &  -0.4$^{\pm  2.2}$\\
\hline
\multirow{2}{0.02\hsize}{Avg.} &  & &   7.8$^{\pm  0.4}$&   3.0$^{\pm  0.4}$&   2.0$^{\pm  0.3}$&   0.8$^{\pm  0.3}$&   1.6$^{\pm  0.3}$\\
 & & &   7.7$^{\pm  0.4}$&   2.8$^{\pm  0.4}$&   1.8$^{\pm  0.3}$&   0.6$^{\pm  0.3}$&   1.4$^{\pm  0.3}$\\
\hline
\multirow{2}{0.02\hsize}{ Kp2} & & &   6.6$^{\pm  0.3}$&   2.4$^{\pm  0.3}$&   1.4$^{\pm  0.3}$&   0.9$^{\pm  0.3}$&   1.2$^{\pm  0.3}$\\
& & &   6.6$^{\pm  0.3}$&   2.4$^{\pm  0.3}$&   1.4$^{\pm  0.3}$&   0.8$^{\pm  0.3}$&   1.1$^{\pm  0.3}$\\
\hline

\end{tabular}\caption{Dust-correlated emissivity in the 5 \emph{WMAP} bands from a C--C analysis with the FDS model 8 dust prediction at 94~GHz.  For each region, the first row gives the result from a simultaneous fit of the three foreground components, while the second row gives that for a fit of two foregrounds to data with the H$\alpha$ (free-free) subtracted assuming $T_e=4000$~K.  Also shown are the FDS8 predicted dust emission intensity range for each region as well as the SFD dust temperature.  The five dust-dominated regions are highlighted in bold face.  Full sky fits, outside the Kp2 cut, are also shown;  here, only a diagonal approximation to the covariance matrix is used, and the uncertainties are determined using simulations. \label{tab:anomalous_dust}} \end{table*}


\subsection{Synchrotron emission}
\label{sec:synch}
The 408~MHz all-sky map is used as the basic
synchrotron template for comparison with the \emph{WMAP} data. The
fits between the 408~MHz and the \emph{WMAP} maps (bands K, Ka and Q)
are given in the upper part of Table~\ref{tab:synch}. The spectral
index of synchrotron emission between the \emph{WMAP} bands can be
derived from the 408~MHz-correlated signal at each \emph{WMAP}
frequency and is given in the bottom part of
Table~\ref{tab:synch}. Note that the implied spectral index can be
misleading where the synchrotron is not detected at an amplitude
higher than its error bar (upper part of Table~\ref{tab:synch}). It
can be seen that, except of region 11 at K-band, the results for
individual regions are not significant at the $2\sigma$
level. Nevertheless, these spectral index values are somewhat steeper
than those calculated between GHz frequencies and the \emph{WMAP}
frequencies.  This is likely to be due to the effect of spectral
ageing of the CR electrons which produce the synchrotron emission in
the Galactic magnetic field. The average of the five regions is
significant at K-band with $\beta=-3.18$ and the full-sky value
(outside of the Kp2 cut) is $-3.01$. The uncertainties shown for the Kp2 fits are based on
simulations and are larger than might be expected since they use a
diagonal approximation to the full covariance matrix M in equation~(4)
(see Appendix~\ref{appendix_cc} for details).



%
\begin{table}\begin{centering}\begin{tabular}{cccc}
\hline
 & \multicolumn{3}{c}{Synchrotron fit amplitudes ($\mu$K K$^{-1}$)} \\
Field & $\frac{K}{408}$ & $\frac{Ka}{408}$ & $\frac{Q}{408}$ 
\\
\hline
11&$ 4.28^{\pm  1.76}$&$-0.02^{\pm  1.11}$&$-0.65^{\pm  0.77}$\\
12&$ 2.94^{\pm  1.64}$&$ 0.55^{\pm  0.63}$&$ 0.13^{\pm  0.26}$\\
13&$ 2.77^{\pm  2.02}$&$ 0.73^{\pm  0.83}$&$ 0.50^{\pm  0.40}$\\
14&$ 1.60^{\pm  1.54}$&$-0.36^{\pm  0.63}$&$-0.49^{\pm  0.31}$\\
15&$ 2.97^{\pm  2.78}$&$-0.82^{\pm  1.51}$&$-0.17^{\pm  0.91}$\\
\hline
{Avg.} &$ 2.82^{\pm  0.82}$&$ 0.15^{\pm  0.36}$&$-0.04^{\pm  0.17}$\\
\hline
{ Kp2} & $  5.56\pm 0.79$& $  1.46\pm 0.78$& $  0.50\pm 0.77$\\
\hline \hline
 & \multicolumn{3}{c}{Synchrotron spectral index $\beta$} \\
Field & $\frac{K}{408}$ & $\frac{Ka}{408}$ & $\frac{Q}{408}$ 
\\
\hline
11&$-3.07^{+ 0.09}_{- 0.13}$&$<-2.97$&$<-3.03$\\
12&$-3.17^{+ 0.11}_{- 0.20}$&$-3.28^{+ 0.17}$&$-3.44^{+ 0.23}$\\
13&$-3.18^{+ 0.14}_{- 0.32}$&$-3.22^{+ 0.17}$&$-3.15^{+ 0.13}_{- 0.35}$\\
14&$-3.32^{+ 0.17}_{- 0.81}$&$<-3.17$&$<-3.47$\\
15&$-3.16^{+ 0.16}_{- 0.69}$&$<-2.97$&$<-2.89$\\
\hline
{Avg.} &$-3.18^{+ 0.06}_{- 0.09}$&$-3.58^{+ 0.28}$&$<-3.27$\\
\hline
{ Kp2} &$-3.01^{+ 0.03}_{- 0.04}$&$-3.06^{+ 0.10}_{- 0.18}$&$-3.15^{+ 0.20}$\\
\hline

\end{tabular}\caption{Synchrotron fits between 408MHz and the \emph{WMAP} K, Ka, and Q bands.  In the lower half of the table, the fit value is converted to a spectral index $\beta$ (T$_b\propto\nu^\beta$), as are that value plus and minus the $1\sigma$ error bars (where positive).  Where the fit amplitude is less than zero, the $2\sigma$ upper limit on the index is shown instead ($<2\sigma$).}
\label{tab:synch}
\end{centering}\end{table}


\section{Discussion}
\label{sec:discussion}

\subsection{Free-free emission}
\label{sec:freefree_discussion}

Free-free emission is the weakest foreground component at \emph{WMAP}
frequencies for intermediate and high Galactic latitudes. By selecting
five \ha-dominated regions, we have been able to quantify the
\ha-correlated free-free emission in these regions as tabulated in
section~\ref{sec:free-free} (Table~\ref{tab:ff_ratios}).

We note that the current analysis is for intermediate and high
latitude H{\sc ii} regions which are therefore associated with the
local spiral arm such as the Gould Belt system and the Gum Nebula;
they most likely lie at $|z| \ltsim 200$~pc from the Galactic
plane. This class of H{\sc ii} regions is different from the more
compact regions confined to the Galactic plane with a width of $z
\approx 60$~pc (Paladini et al. 2003, 2004). As mentioned in
section~\ref{sec:ha_template}, these H{\sc ii} regions have a mean
$T_{e}=7200 \pm 1200$~K
compared to the value of around $4000-5000$~K found in the present
study. Note that for the brightest region (5) where $T_{e}$ is most
accurately determined, $T_{e} \sim 6500$~K. This discrepancy has an
unclear origin, especially given that an identical result is
determined for the entire high-latitude sky as defined  by the
\emph{WMAP} Kp2 mask. It may be indicative of problems associated with
the \ha\ template itself, or in the conversion of \ha\ flux to the
free-free brightness temperature. However, variations by a factor of
$\sim$2 are also seen from region to region.

It is of interest to note that the electron temperature derived from
radio recombination line studies of extended H{\sc ii} regions such as
the Gum Nebula
have an average value of $~7000$~K (Woermann, Gaylard \& Otrupcek
2000). A study of diffuse foregrounds in the
  \emph{COBE}-DMR at $7\deg$ angular scales (Banday et al.~2003) found
  that \ha~correlations with the DDD \ha~template were more or less
  consistent with $T_{e}\sim 7000$~K for $|b| > 15\deg$. For $|b| >
  30\deg$, lower values were preferred but with larger error
  bars. Banday et al.~(2003) also analysed 19~GHz data with a $3\deg$
  beam, which favoured lower values even for $|b|>30\deg$. These
  results suggest that the discrepancy may be scale-dependent and
  therefore might be related to the different beam shapes of the WHAM
  and SHASSA \ha~surveys for angular scales comparable to the beam
  size ($\sim 1\deg$).


\subsection{Dust emission}
\label{sec:dust_discussion}

\begin{figure*}
\begin{center}
\includegraphics[width=0.75\textwidth]{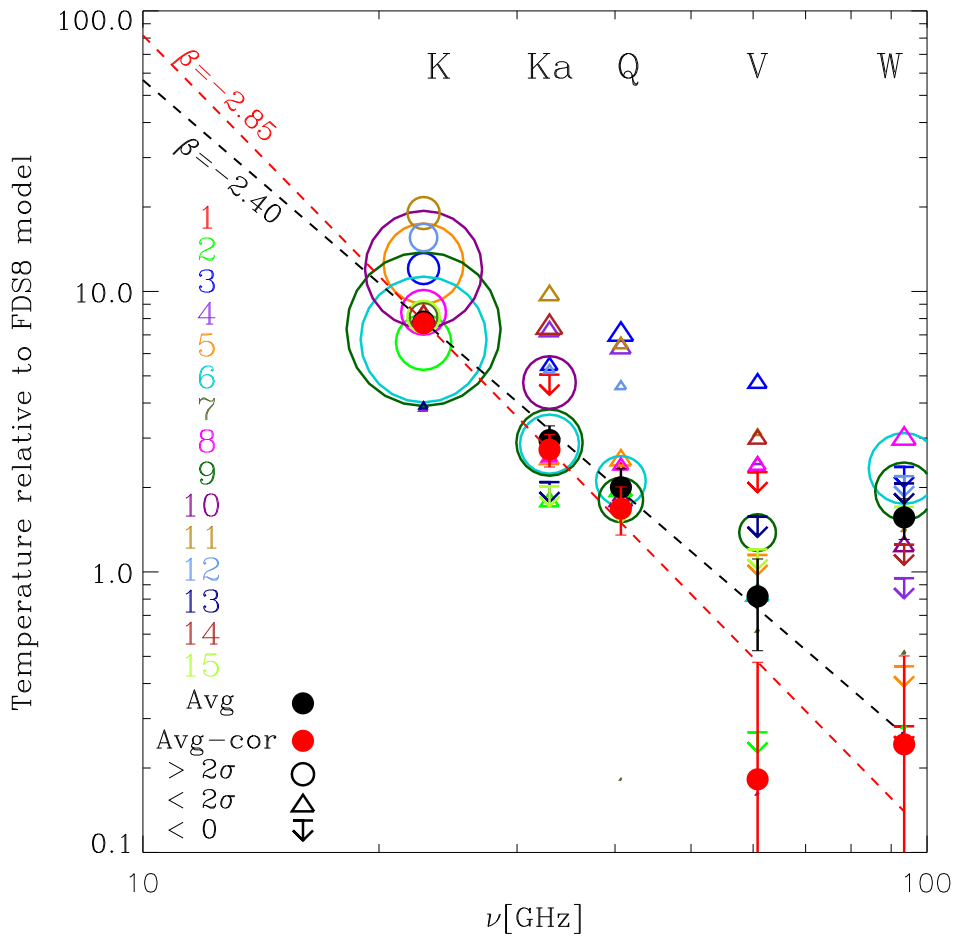}
\caption{Summary of dust emissivities (antenna temperature units,
relative to FDS8 at 94~GHz) from C--C analysis.  Each region emissivity is plotted
with a different colour open symbol whose size is proportional to the
signal-to-noise ratio.  The symbol is a circle if the signal-to-noise
is greater than two or a triangle if it is positive but less than two.
Where the fit value is negative, the $1\sigma$ upper limit is shown as
an arrow downward, and values more than one sigma negative are not
shown (though all values are included in the weighted average).  The
filled black circle shows the weighted average surrounded by its error
bar.  The dashed line represents the best-fit spectral index over K
through Q bands.  Red filled circles and the red dashed line show the
average and fit when corrected for the best-fit thermal dust component
assuming $\beta=+1.7$; see text in \S \ref{sec:dust_discussion}.
\label{fig:dustem}}
\end{center}
\end{figure*}

The dust-correlated emission is the dominant foreground component in
the \emph{WMAP} bands and its spectral properties can be derived for
the individual clouds included in the present study. The spectra of
all fifteen regions are shown in  Fig.~\ref{fig:dustem}. It is
immediately seen that the spectral slopes of each of the clouds over
the range from K- to V-band are quite similar. Also, all the clouds
show a turn-up in emissivity at W-band where thermal emission becomes
dominant.

A further significant result is that the emissivity relative to the
FDS8 prediction varies by a factor of 2 from cloud to cloud.

The average spectral emissivity for the clouds is shown by black
filled  circles in Fig.~\ref{fig:dustem}. The average spectral index
from K-
to Q-band, is $-2.4$, shown as a dashed black line in
Fig.~\ref{fig:dustem}.  The average spectral index in the range
K--Ka-band and Ka--Q-band are $-2.6$ and $-1.9$.   An estimate of
$\beta$ from K-band to higher frequencies depends sensitively upon the
vibrating dust contribution in these bands and requires a knowledge of
its spectral index.  Assuming $\beta=+1.7$,
we use a simple $\chi^2$ test to find the best fit to the data over a
grid of values for the anomalous dust spectral index, the anomalous
dust amplitude, and the vibrating dust amplitude.  The result gives a
spectral index for the anomalous component of -2.85 and shows that the
FDS8 prediction at W-band is underestimated by 30 per cent. Assuming a
steeper thermal index of 2.0 or 2.2 results in an anomalous index of
$-2.75$.  The data minus the vibrating dust fit are shown in red in
Fig.~\ref{fig:dustem} along with the best-fit anomalous power law.

One proposed explanation for the anomalous dust-correlated emission in
the low frequency \emph{WMAP} bands, motivated by its spectral
behaviour, is that it represents a hard synchrotron component,
morphologically different in the \emph{WMAP} bands from the soft
synchrotron component traced by the 408 MHz emission.  This hard
synchrotron emission would correlate with dust in regions of active
star formation.  We find that this anomalous component has a spectral
index from the K- to Q-band of $\beta=-2.85$ when the FDS8 thermal
dust prediction is subtracted, assuming $\beta=+1.7$ (see
Fig.~\ref{fig:dustem}).   We then extrapolate this component to 408
MHz to see how much of this hard emission would be seen at that low
frequency.  If there is no spectral hardening between 408 MHz and
\emph{WMAP}, then a spectral index of $\beta=-2.85$ would imply more
emission at 408 MHz than is observed by a factor of $\gtsim 2$ in many
regions.  If the thermal dust spectrum is steeper, e.g., $\beta=+2.2$,
then (as discussed above) the anomalous index flattens slightly to
$-2.75$, but that still over-predicts the emission at 408 MHz.

We now consider the relevance of the dust temperature to our
results, using as a proxy the SFD98 colour temperature
based on the ratio of the DIRBE 100- and 240-~$\mu$m 
data at a resolution of $1.\!\deg1$.

\begin{figure}
\begin{centering}
\includegraphics[width=0.35\textwidth]{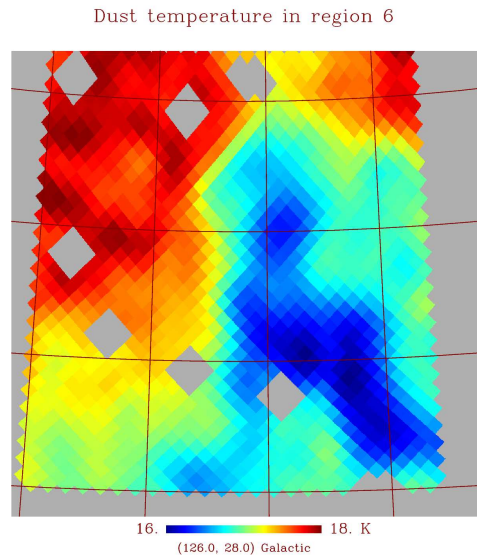}
\caption{SFD dust temperature for region 6, showing that the emissivity, as seen in Fig.~\ref{fig:region_example_maps}, comes from the coldest regions. \label{fig:dust_t_r6}}
\end{centering}
\end{figure}

Comparing the dust temperature to the emissivity in these regions (as
well as over the sky outside the Kp2 cut), it appears that in general,
the strongest anomalous emission (relative to the FDS8 prediction) comes from the
coldest regions.  This is particularly striking in the two dust
regions which have the smallest error bars, 6 and 9, which can be
seen to dominate the averages in Table \ref{tab:anomalous_dust} and
Fig.~\ref{fig:dustem}.  
Fig.~\ref{fig:dust_t_r6} shows the dust temperature in region 6, which
can be compared to the $100~\mu$m SFD98 map shown in Fig.~\ref{fig:region_example_maps}.
The fit amplitudes in these regions
show the lowest cross-correlation with the
low-frequency data, where the emission is not thermal but comes from
the anomalous component. Table
\ref{tab:anomalous_dust} shows that the K-band emission (where
strongly detected) is lowest relative to the FDS8 template in regions 2, 6, and
9, which have the lowest dust temperatures as well.  (Not apparent
from the table, which shows only the dust temperature {\em range}, is
that the emission comes from the coldest parts of the region).

Finkbeiner (2004) has examined the foreground residuals after
subtracting the FDS8 prediction and proposed that an anomalous
dust template could be better constructed 
using FDS8$\times {\rm T}^2$.  By comparing the $\chi^2$
values for template fits for the Kp2 mask
using FDS8$\times {\rm T}^n$ for different
values of $n$, we find that formally the best value for $n$ in the
K-band is $1.6$ and that it drops to zero at the higher
frequencies, as one would expect.  The same exercise repeated on the fifteen
regions, however, gives a large scatter in the preferred value of $n$,
ranging from 0 to over 5 (the limit of the range tested).  

The average dust emissivity among all of the regions is slightly
higher at low frequencies (particularly K-band) than that over the
full sky (outside the Kp2 cut).  Can this be explained by the full sky
fits being driven by the even colder emission near the Galactic Plane
at the anti-centre?  Fits of the FDS8 template to the hemisphere around
the Galactic centre and around the Galactic anti-centre do indeed show
that the fit values around the anti-centre are from 10 to 30 per cent lower
than fits around the Galactic centre, depending on the band.  These
differences appear to confirm the indication in the smaller regions
that the emissivity of the anomalous component is lower from colder
dust.

\subsection{Synchrotron emission}
\label{sec:synch_discussion}

As in the case of free-free emission the synchrotron
emission, defined as the 408~MHz-correlated emission, is weaker than
the dust-correlated emission at \emph{WMAP} frequencies ($20-90$~GHz)
over most of the sky. Except for the strongest synchrotron feature, the North Polar Spur (region 11), the 408-MHz correlations are marginal, even at K-band. Nevertheless the average
spectral index of the 5 synchrotron-selected fields
(Table~\ref{tab:synch}) indicates an increasing slope relative to 408~MHz
of $\beta=-3.18$ to $<-3.27$ from K to Q-band.  The full-sky fits
(Kp2) indicates $\beta=-3.01$ from 408~MHz to 23~GHz. At higher WMAP
frequencies, the fits are not significant. The errors are probably too
conservative due to the diagonal covariance assumption used in the Kp2
fit (see Appendix \ref{appendix_cc}). Nevertheless, the results are in
good agreement with the \emph{WMAP} values reported in Bennett et
al.~(2003b).

Using low frequency maps at GHz frequencies (408~MHz,
1420~MHz, 2326~MHz from Haslam et al. (1982), Reich \& Reich (1986)
and Jonas et al. (1998), respectively) we find that the average
spectral index for the 5 maps in the GHz range is $-2.91$.  This
steepening with frequency is indicative of ageing of the relativistic
electrons in these fields. This can be compared with the results of
the Cosmosomas experiment which found $\beta~-3.16$ from 408~MHz to
13~GHz for $|b|>30\deg$ (Fern\'{a}ndez-Cerezo et al.~2006). Our limited data does not
show evidence for strong variation of the synchrotron spectral index
at \emph{WMAP} frequencies from field-to-field. The discrepancy
between the average of the regions and the Kp2 cut is likely to be due
to the dominance of the North Polar Spur which is known to have a
steeper index relative to the sky average.


\section{Comparison with \emph{WMAP3}}
\label{sec:wmap3}

Whilst this paper was being completed, the \emph{WMAP} team released
their 3-year results to the community.  For this new analysis, several
semi-independent studies of the foreground contamination of the data
were undertaken (Hinshaw et al.~2006).  For the purposes of gaining
physical insight into the nature of the Galactic foregrounds, a
maximum-entropy (MEM) technique was applied. However, for the purposes
of cleaning the data for cosmological studies, a template subtraction
method was adopted. As with the first year analysis, the F03 \ha\
template was employed as a tracer of free-free emission and the FDS8
model normalised at 94 GHz used for thermal dust emission. For
synchrotron emission an internally generated template, comprising the
difference of the K- and Ka-bands, was constructed. There are several
aspects of these foreground results that merit comment in the present
paper.

Hinshaw et al. (2006) have determined a free-free to \ha\ ratio of
$\sim$6.5~$\mu$K~R$^{-1}$ based on fits to the F03 template. This is
completely consistent with the mean values derived from the 5
free-free regions in this paper, particularly after adjusting the
\emph{WMAP} coefficient 20-30 per cent upwards to compensate for the
increased amplitude of their $f_{d}=0.5$ dust corrected \ha\ template.
The MEM analysis finds a slightly higher ratio
$\sim$8~$\mu$K~R$^{-1}$, but also considerable variation (by a factor
of $\sim$2) depending on location, as we have also found.


They adopted the difference between the observed K- and Ka-band
emission as a tracer of synchrotron emission was intended to
compensate for the problem with assuming a fixed full-sky spectral
index in order to extrapolate the Haslam 408~MHz sky map to
\emph{WMAP} frequencies.  Such a procedure is clearly in contradiction
with the spectral index studies of Reich \& Reich (1988) between 408
and 1420~MHz, which showed large variations of
spectral index across the sky. Of course, utilising the K-Ka map as a
foreground correction template is, to some extent, independent of
whether the dominant foreground contribution is due to synchrotron,
anomalous dust, or a combination thereof. By utilising what Hansen et
al. (2006) have referred to as an internal template, it is likely that
the synchrotron morphology is well traced over the frequencies of
interest.  Moreover, fitting this template to the remaining sky maps
with a global scale factor per frequency is likely to be quite
accurate, even in the presence of modest departures from a single
spectral index.  This treatment does not contradict our own studies,
since our intention is to study the variations in spectral behaviour
over the sky relative to the 408~MHz survey
(Section~\ref{sec:synch_discussion}).

Maintaining a consistent approach to their treatment of the first year
data, the \emph{WMAP} team have not attempted to directly address the
issue of the anomalous dust correlated component. Rather, the MEM
solutions were allowed only to produce what may be interpreted as a
combined synchrotron/anomalous dust solution at each frequency, with
no attempt made to disentangle the two components. Hinshaw et
al. (2006) comment that it is not possible, using the \emph{WMAP} data
alone, to distinguish between anomalous dust emission and flatter
spectrum synchrotron emission that is well correlated with dusty
star-forming regions. This is particularly true given that a putative
spinning dust component can exhibit a similar spectral shape to
synchrotron emission over the 20 - 40 GHz frequency range. However,
Page et al. (2006) in their foreground modelling efforts for the
polarisation analysis, determine a high-polarisation fraction
component of the synchrotron emission that is well correlated with the
Haslam template, and a low-polarisation component with a dust-like
morphology. We argue that this is indicative of a spinning dust
component. Furthermore, an unexpected flattening of the spectral index
for the polarised synchrotron emission was also found. It is important
for understanding the foreground polarisation where the percentage
polarisation is very different for each component. 


Finally, we note that the \emph{WMAP} team impose a constraint on
their thermal dust template fits that the derived dust coefficients
must have a spectral index of $\sim$2, rather than the value of 2.2
used in the first year analysis, or the value of 1.7 predicted by the
FDS8 model. This makes only a very minor difference to
the spinning dust spectrum because thermal dust emission is negligible
at in the lowest \emph{WMAP} bands.


\section{Conclusions}
\label{sec:conclusions}

In our study of the free-free, dust and synchrotron foreground
components in the \emph{WMAP} data we have chosen a selection of
fields which are intended to have minimal cross-contamination from other
components. Each of the 3 components has been quantified in terms of a
mean value of the emissivity in each of the 5 \emph{WMAP} bands. 

\begin{figure*}
\begin{center}
\includegraphics[width=0.75\textwidth,angle=0]{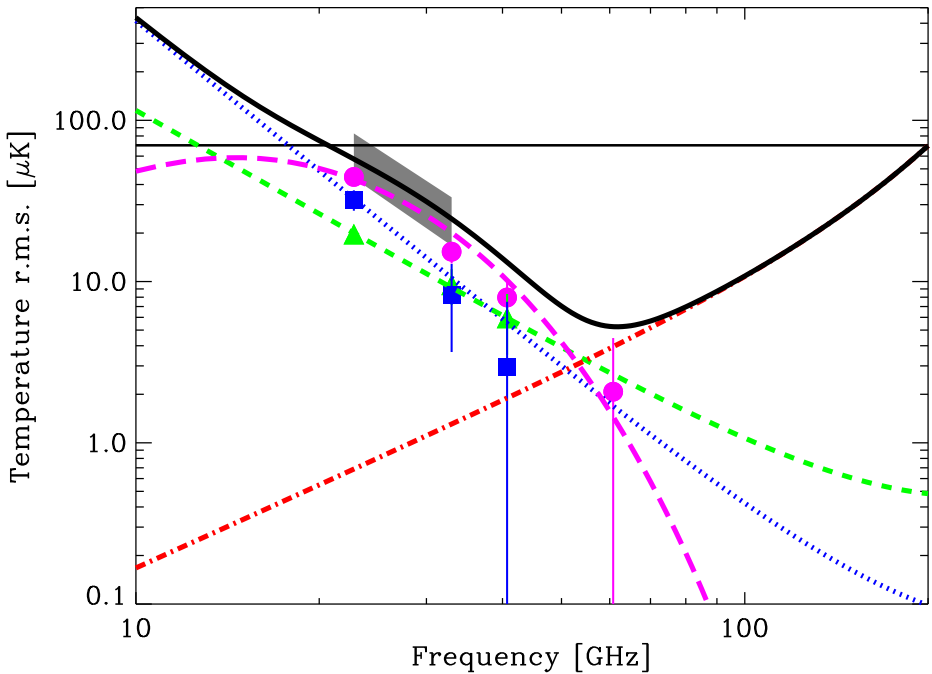}
\caption{R.m.s. fluctuations spectrum (thermodynamic temperature units) of foreground components at $1\deg$ resolution. The symbols are the Kp2 solutions from the C--C analysis for synchrotron ({\it blue squares}), free-free ({\it green triangles}) and anomalous dust-correlated emission ({\it magenta circles}). The curves represent synchrotron for $\beta=-3.1$ ({\it blue dotted line}), free-free for $\beta=-2.14$ ({\it green short dashed line}), vibrational dust with $\beta=+1.7$ ({\it red dot-dashed line}), Draine \& Lazarian spinning dust model ({\it long dashed magenta line}) and all the foreground models combined in quadrature ({\it thick black line}). The CMB fluctuations are shown at $70~\mu$K r.m.s ({\it thin black line}). The grey region shows the variation of dust emissivity between K- and Ka-bands in regions where significant ($> 2\sigma$) dust-correlated emission was detected.
\label{fig:spectrum_rms}}
\end{center}
\end{figure*}

Fig.~\ref{fig:spectrum_rms} shows the emission, in thermodynamic
units, as expected using the values of emissivity we have determined
in conjunction with the \ha, 100~$\mu$m dust and the 408~MHz templates
outside the Kp2 mask. These are calculated using r.m.s. values of
5.9~K, 2.6~R and $6.8~\mu$K for 408 MHz, \ha, and FDS8 model at
94~GHz, respectively. The data points are the Kp2 solutions from the
C-C  analysis. The curves are foreground models; synchrotron with
$\beta=-3.1$ normalised to K-band, free-free with $\beta=-2.14$ for
$T_{e}=4000$~K and vibrational dust emission for $\beta=+1.7$
normalised to W-band. The magenta curve is a spinning dust model from
Draine \& Lazarian (1998a,b)\footnote{Spinning dust models were
downloaded from: http://www.astro.princeton.edu/\,$\tilde{}$
draine/dust/dust.html/.}, scaled to fit the data points from K- to
Q-band. The curves plotted in Fig.~\ref{fig:spectrum_rms} are
therefore not strictly ``best-fits'' to these data points and are
plotted to depict the approximate amplitude and spectral dependencies
of the 4 Galactic components at high latitudes (outside the Kp2 cut). The dominance of the
dust emission is evident. Also the similarity of the dust and the
synchrotron spectrum at \emph{WMAP} frequencies is superficially
evident; these may be separated by lower frequency ($5-15$~GHz) data
as shown for example by de Oliveira-Costa et al.~(2004) and Watson et
al.~(2005). We note that dust (anomalous and thermal) is the dominant
foreground over the \emph{WMAP} and Planck bands. The thick black
curve in Fig.~\ref{fig:spectrum_rms} is the total of the 4 model
curves, combined in quadrature, corresponding to the approximate total
foreground r.m.s. level; the minimum foreground contamination of the
CMB is at $\sim 70$~GHz for total intensity. The
integrated foreground spectrum is relatively simple when sampled
sparsely in frequency (e.g. \emph{WMAP}). This is why the \emph{WMAP}
team find that the spectrum from 408~MHz to V-band is well-fitted by a
simple power-law, although it says little in itself about the
underlying foreground components.

One of the most interesting results of the present study is the
variation in the emissivity of the dust-correlated emission in the
 \emph{WMAP} bands, typically by a factor of $\sim 2$. The grey shaded
region of Fig.~\ref{fig:spectrum_rms} indicates the variation of the
dust-correlated component between K- and Ka-bands from the C-C
analysis of 15 regions where a significant ($>2\sigma$) detection was
found. There is clearly considerable variation in dust emissivity at
K- and Ka-bands. Fig.~\ref{fig:spectrum_rms} also
shows that the average emissivity in the regions is higher than
typical value seen outside the Kp2 cut (see
Table~\ref{tab:anomalous_dust}). The effect on producing an anomalous
dust template is profound because it is the dominant foreground
between 20 and 60~GHz. If the anomalous dust emission is polarised at
$\sim 10$~per cent this could approach the synchrotron emission as an
important polarised foreground in the $20-60$~GHz range. These
considerations are important for cleaning CMB maps from missions such
as \emph{WMAP} and Planck as well and high sensitivity ground-based
experiments such as
CLOVER\footnote{http://www-astro.physics.ox.ac.uk/research/expcosmology/groupclover.html}
and QUIET\footnote{http://quiet.uchicago.edu/}.

The \emph{WMAP} team have instigated some debate over the origin of
the anomalous dust correlated component, and have preferred an
interpretation in terms of a hard synchrotron contribution from
star-forming regions that are strongly associated with dust. The 5
synchrotron regions selected in this paper are dominated by well-known
structures on the sky away from star-forming regions.  As such, it
might be expected that the derived synchrotron indices would be steep,
and this is indeed the case.  It is also expected that cross-talk with
other physical components should be minimised. An
important results is that anomalous emission was detected in 11 out of
the 15 regions of studied here.  We consider this to be strong
evidence against a synchrotron origin for the anomalous component,
although the exact nature of the dust emission mechanism requires
lower frequency measurements to elucidate the detailed spectral
behaviour. Certainly, the anomalous emission constitutes the dominant
foreground component over the 20 -- 60 GHz frequency range.

Further work is required to understand the origin of this variation in
dust emissivity by using other physical properties of dust such as its
size and temperature. New data in the critical radio frequency range
$\sim 5-15$~GHz will be vital for a clearer definition of the
anomalous dust spectrum. Polarisation data will be particularly
important for understanding the physical mechanism that produces the
anomalous emission which is expected to be polarised at different
levels (e.g. Draine \& Lazarian 1999). For
example, spinning dust emission is expected to be only weakly
polarised, whereas the synchrotron emission is known to be highly
polarised.


\section*{ACKNOWLEDGMENTS}

CD warmly thanks Barbara and Stanley Rawn Jr for funding a research
scholarship at the California Institute of Technology. Some of the
results in this paper have been derived using the HEALPix package
(G\'{o}rski et al.~2005). We acknowledge the use of the Legacy Archive
for Microwave Background Data Analysis (LAMBDA). Support for LAMBDA is
provided by the NASA Office of Space Science. The Wisconsin H-Alpha
Mapper (WHAM) is funded by the National Science Foundation. We thank
Bruce Draine for making his spinning dust models available.


\bibliographystyle{mn2e}


\bsp 


\appendix

\section{Aliasing of foregrounds due to CMB subtraction}
\label{sec:aliasing}

  Some studies of the \emph{WMAP} data remove the ILC map before
performing the correlation analysis in order to minimise the impact of
the CMB structure on that analysis. However, this introduces a new
problem. The ILC is constructed as a linear combination of the five
\emph{WMAP} frequency bands in such a way that i) the CMB signal is
conserved; ii) the variance of the final map is minimised. The latter
condition does \emph{not} guarantee the absence of residual
foregrounds from the ILC. Therefore, when one subtracts the ILC from
the individual \emph{WMAP} frequency maps the foreground contribution
is altered by an amount depending on the residuals in the ILC
map. Since we do not know a priori the actual properties of the
foregrounds present in the data, it is not possible to absolutely
correct for the effect of this 'aliasing' of foregrounds from the ILC
into the individual frequency maps. However, as shown in Eriksen et
al. (2004), the likely residual level can be predicted under various
assumptions about the foreground spectral behaviour. This at least
gives some insight into the impact of the ILC subtraction on derived
foreground properties.

For the high latitude fields considered in this paper, the 5
coefficients that define the ILC map for each frequency band are
constant.\footnote{Bennett et al. (2003b) also derive coefficients in
12 regions of the Galactic plane.} This allows us to calculate the
effect of the ILC subtraction, for a given spectral index, both in
terms of relative amplitude of the aliased signal as well as the
effect on the derived spectral index.

The numbers in Table~\ref{tab:ilc_fg_resids} represent the fraction of
foreground signal present in the corrected frequency map in antenna
temperature units, assuming a specific spectral behaviour for that
foreground.  The CMB correction due to the ILC is given
by 0.109K$-$0.684Ka$-$0.096Q$+$1.921V$-$0.250W.  For these
coefficients, it turns out that the effect can be relatively
small for some of the foregrounds and frequencies of the \emph{WMAP}
data.  
For example, a
synchrotron component, with $\beta=-3.1$ is increased at the 3 per
cent level at K-band or 10 per cent at Ka-band. For synchrotron, the
strong aliasing at high frequencies is not important since the actual
foreground level is well below that of the other components. For \ha,
with a flatter spectral index, the effect is $\sim$1
per cent lower at K-band and $\sim$15 per cent at W-band. 
It appears that the K-, Ka- and Q-band fits are reliable tracers of the
synchrotron and free-free spectral indices up to a modest correction factor.
The anomalous emission appears to have a relatively steep
spectral index similar to that of synchrotron 
thus we might expect aliasing at similar levels to the synchrotron
emission for $\beta=-2.85$. This may not be the case if 
the spinning dust models of Draine \& Lazarian are
correct models of the emission mechanism.

The aliasing effect is most strongly seen in W-band for the
vibrational dust component. For $\beta=+2.0$, the dust component will
be {\it reduced} by as much as 40 per cent. This is enough to change the effective
spectral index of this component by a significant amount; a true
spectral index of $+1.7$ increases to $+1.9$ after ILC
subtraction. Furthermore, the dust correlation coefficients
at low frequency have oversubtracted thermal dust contributions.

It is our contention that making an ILC subtraction before foreground
analysis introduces difficulties of interpretation that may invalidate
conclusions unless the effect of foreground aliasing is handled
correctly. 

\begin{table}
\begin{tabular}{cccccc}
\hline
$\beta$   & K      & Ka       & Q        & V        & W \\
\hline

\multicolumn{6}{c}{\it Free-free} \\
  -2.0&           0.98&     0.96&     0.93&     0.86&     0.71\\
  -2.1&           0.99&     0.98&     0.97&     0.93&     0.84\\
  -2.2&           1.00&     1.00&     1.00&     0.99&     0.99\\
\hline
\multicolumn{6}{c}{\it Thermal dust}\\
   1.7&          -5.19&    -2.25&    -1.24&    -0.08&     0.54\\
   2.0&          -6.97&    -2.75&    -1.43&    -0.03&     0.62\\
   2.2&          -8.34&    -3.08&    -1.54&     0.00&     0.66\\
\hline
\multicolumn{6}{c}{\it Anomalous dust}\\
  -2.5&           1.02&     1.05&     1.08&     1.21&     1.53\\
   CNM&           1.09&     1.08&     1.09&     1.24&     3.59\\
   WNM&           1.28&     1.41&     1.69&     5.41&   436.00$^*$\\
   WIM&           1.28&     1.32&     1.46&     3.01&    62.10\\
\hline
\multicolumn{6}{c}{\it Synchrotron} \\
  -2.7&           1.03&     1.07&     1.12&     1.35&     1.98\\
  -2.9&           1.03&     1.09&     1.16&     1.49&     2.51\\
  -3.1&           1.03&     1.10&     1.19&     1.62&     3.09\\
\hline
\end{tabular}
\caption{
The fraction of foreground signal present in the corrected frequency
map in antenna temperature units, assuming a specific spectral
behaviour for that foreground. For example,. the Q-band contains 119 per cent of the
synchrotron contribution expected if the synchrotron emission has a
spectral index of $\beta=-3.1$. Therefore, in the absence of
cross-talk with other foreground components, a synchrotron template
fit to the Q-band should be downweighted by a factor of 1.19.  
For anomalous dust, we also tabulate the expected aliasing signal
for the Cold Neutral Medium (CNM), Warm Neutral Medium (WNM),
and Warm Ionised Medium (WIM) from models due to Draine \& Lazarian
(1998a,b).
$^*$The W-band aliasing factor is large, but since the WNM spectrum falls off very
rapidly this factor results only in physical amplitudes closer to the other channels.
\label{tab:ilc_fg_resids}}
\end{table}


\section{The reliability of T--T versus C--C fits}\label{appendix_cc}

The T--T method is a straightforward linear fit of two datasets and has
two basic drawbacks: firstly, only one template can be compared to the
data, and secondly, only independent pixel noise (effectively, a
diagonal covariance matrix) can be taken into account.  The fifteen
regions were chosen such that one component is dominant in an effort
to get around the first problem.  The ILC can be subtracted from the
data in order to deal with the second.  The C--C method is more
complicated but allows simultaneous fitting of multiple templates with
a full signal covariance matrix.

We use simulations with foreground components added at known
amplitudes to test how well these different fitting methods recover
the input.  A set of 1000 simulations are created at the resolution of
the
\emph{WMAP} data products, i.e. at HEALPix of $N_{\textrm
side}=512$, using the beam width and noise properties of the
respective channels.  The foregrounds are added at the theoretical
levels for free-free (assuming an electron temperature of T$_e$=8000~K)
and for synchrotron (assuming a spectral index of $\beta=-3.0$), while
the dust is added at the levels found by Table~3 of Bennett et
al. (2003b), approximating both thermal and anomalous contributions.
These maps are then smoothed (via convolution in harmonic space) to a
common resolution of $1\degr$ and downgraded to $N_{\textrm
side}=128$, as are the data.


These simulations can be used to test how much cross-talk still
affects fits using only one template in the small regions chosen in an
attempt to minimize such problems.  In other words, does the existence
of small amounts of additional dust and synchrotron emission in the
H$_\alpha$-dominated regions affect the individual fits using the
H$_\alpha$ template alone?  Looking at the elements of the fit matrix
$A$ (defined in \S
\ref{sec:cc_method}) gives an indication of how the different
templates correlate with each other.  This is shown for a few regions
in Table \ref{tab:cross_cor}.  But analysis of simulations is needed
to quantify the effects on the fit results.

First, we compare fits to data where the ILC estimate of the CMB
component is subtracted, which is necessary in the T--T method but which
introduces the small aliasing effects discussed in Appendix
\ref{sec:aliasing}.  Table \ref{tab:compare_tt_cc} columns labeled
``1to3'' give results where one template is fit to simulations with
three foregrounds; these show that the H$_\alpha$ estimates {\it are}
overestimated due to the other emission, even in regions dominated by
H$_\alpha$ emission.  The amount depends on the region, but some
regions are overestimated by as much as 30\%.  This affects both
methods when only one template is used.  But where the C--C method is
given the three templates to fit simultaneously, the results are then
roughly consistent with the input, as seen in columns labeled
``3to3''.  The C--C and T--T methods also both give roughly correct
results when only one foreground component is present in the data, as
seen in the ``1to1'' columns.

The bias due to the ILC subtraction is clear in Table
\ref{tab:compare_tt_cc} in the ``1to1'' columns or the first ``3to3''
column.  As expected for an index of $\beta=-2.1$, the free-free
emission is underestimated by $\sim 1\%$.  Likewise, the synchrotron
is overestimated by approximately 3\%.  The dust is underestimated by
$\sim6\%$, as expected for the mixture of thermal and anomalous dust
represented by the input amplitudes, taken from the \emph{WMAP} best
fits (Bennett et al. 2003b, Table 3.)


We conclude from the above and from the information in Table
\ref{tab:compare_tt_cc} that the only unbiased estimate of the
foreground fit amplitudes comes from the full C--C method using the
three template simultaneously and the CMB signal plus noise covariance
matrix against the raw data.  The bias introduced by the
ILC-subtraction is relatively small (if the foregrounds follow roughly
the expected spectral behavior), but the cross-talk among templates
prevents the T--T method from being accurate enough in most cases.  Only
the C--C method with the full covariance, shown in the ``3to3$^\dagger$''
column of Table \ref{tab:compare_tt_cc} gives unbiased results.  

In addition to template fitting in small regions, we have also tested
the C--C method on the full sky outside the Kp2 mask.  Only at very low
resolution ($N_{\rm side} <= 32$) can we use the full pixel-to-pixel
covariance matrix, since higher resolution requires storing and
inverting a matrix of $\sim$2GB, which is computationally prohibitive.
For our analysis at $N_{\rm side} =128$, we simply use the diagonal of
the full matrix.  Using the same set of simulations described above,
we can test how well this method works at recovering the input
template amplitudes.  Unlike the case where the full matrix is used,
the diagonal of the matrix $\bf{A}$ does not give an accurate estimate
of the uncertainty in the result, so these simulations are also
necessary to quantify the distributions.  We find that the method is
unbiased, returning the correct mean for each template amplitude.  The
uncertainties are larger than they would be if we could use the full
covariance matrix; using the diagonal makes the method much less
accurate.  This accounts for the fact that the errors shown in Tables
\ref{tab:ff_ratios} through \ref{tab:synch} for the averages over all
the regions are comparable to those given for the Kp2 fits despite the
fact that if the same method could be used in both cases, the Kp2 fits
would have smaller uncertainties.

\begin{table}\begin{tabular}{ccccc}\hline
  & FDS dust & DDD H$_\alpha$ & Haslam & Const.\\
\hline\multicolumn{4}{c}{Region 5}\\
FDS dust  &  1.00 &  0.23 &  0.59 &  0.62\\
DDD H$_\alpha$  &  0.23 &  1.00 &  0.20 &  0.14\\
Haslam  &  0.59 &  0.20 &  1.00 &  0.82\\
Const.  &  0.62 &  0.14 &  0.82 &  1.00\\
\hline\multicolumn{4}{c}{Region 9}\\
FDS dust  &  1.00 &  0.26 &  0.33 &  0.30\\
DDD H$_\alpha$  &  0.26 &  1.00 &  0.42 &  0.44\\
Haslam  &  0.33 &  0.42 &  1.00 &  0.91\\
Const.  &  0.30 &  0.44 &  0.91 &  1.00\\
\hline\multicolumn{4}{c}{Region 14}\\
FDS dust  &  1.00 &  0.34 &  0.51 &  0.61\\
DDD H$_\alpha$  &  0.34 &  1.00 &  0.29 &  0.33\\
Haslam  &  0.51 &  0.29 &  1.00 &  0.73\\
Const.  &  0.61 &  0.33 &  0.73 &  1.00\\
\hline
\end{tabular}
\caption{Fit matrix showing effectively the amount of cross-correlation among the different templates.  Each value is $t^T_i M^{-1} t_j /\sqrt{ (t^T_i M^{-1} t_i)(t^T_j M^{-1} t_j)}$.  The regions were chosen as those where the dominant component had the smallest error bar.    \label{tab:cross_cor} }
\end{table}


\begin{table*}
\begin{tabular}{cccccccc}
\hline
\multicolumn{8}{c}{\bf Average fit amplitude as fraction of true} \\ 
\hline
& \multicolumn{2}{l}{TT...} & \multicolumn{5}{l}{CC...} \\
& 1to3 & 1to1 & 1to3 & 1to1 & 3to3 & 1to3$^\dagger$ & 3to3$^\dagger$ \\
\hline
\multicolumn{8}{c}{free-free (at $11.4~\mu$K~R$^{-1}$)}\\
       1 &   1.02$\pm   0.11$ &   0.99$\pm   0.11$ &   1.03$\pm   0.10(  0.10)$ &   0.99$\pm   0.10(  0.10)$ &   0.99$\pm   0.10(  0.11)$ &   1.12$\pm   0.33(  0.34)$ &   1.00$\pm   0.34(  0.35)$\\
       2 &   1.08$\pm   0.09$ &   0.99$\pm   0.09$ &   1.18$\pm   0.07(  0.08)$ &   0.99$\pm   0.07(  0.08)$ &   0.99$\pm   0.08(  0.09)$ &   1.13$\pm   0.13(  0.16)$ &   1.00$\pm   0.14(  0.16)$\\
       3 &   1.21$\pm   0.07$ &   0.99$\pm   0.07$ &   1.21$\pm   0.07(  0.07)$ &   0.99$\pm   0.07(  0.07)$ &   0.99$\pm   0.09(  0.09)$ &   1.09$\pm   0.16(  0.17)$ &   0.99$\pm   0.17(  0.18)$\\
       4 &   1.14$\pm   0.06$ &   0.99$\pm   0.06$ &   1.14$\pm   0.05(  0.06)$ &   0.99$\pm   0.05(  0.06)$ &   0.99$\pm   0.07(  0.08)$ &   1.08$\pm   0.16(  0.17)$ &   1.01$\pm   0.17(  0.18)$\\
       5 &   1.32$\pm   0.03$ &   0.99$\pm   0.03$ &   1.25$\pm   0.03(  0.03)$ &   0.99$\pm   0.03(  0.03)$ &   0.99$\pm   0.03(  0.03)$ &   1.04$\pm   0.06(  0.06)$ &   1.00$\pm   0.06(  0.06)$\\
\hline
\multicolumn{8}{c}{dust (at $6.3~\mu$K~$\mu$K$^{-1}_{FDS8}$)}\\
       6 &   1.06$\pm   0.03$ & -  &   1.07$\pm   0.03(  0.03)$ & -  &   0.94$\pm   0.03(  0.03)$ &   1.05$\pm   0.11(  0.11)$ &   1.00$\pm   0.11(  0.11)$\\
       7 &   1.05$\pm   0.22$ & -  &   1.06$\pm   0.22(  0.22)$ & -  &   0.94$\pm   0.23(  0.25)$ &   1.13$\pm   0.59(  0.60)$ &   1.00$\pm   0.59(  0.61)$\\
       8 &   0.87$\pm   0.15$ & -  &   0.89$\pm   0.15(  0.16)$ & -  &   0.93$\pm   0.15(  0.16)$ &   0.98$\pm   0.36(  0.37)$ &   1.01$\pm   0.36(  0.37)$\\
       9 &   0.98$\pm   0.05$ & -  &   0.98$\pm   0.05(  0.05)$ & -  &   0.94$\pm   0.05(  0.06)$ &   1.03$\pm   0.10(  0.10)$ &   0.99$\pm   0.10(  0.10)$\\
      10 &   1.03$\pm   0.06$ & -  &   1.03$\pm   0.06(  0.06)$ & -  &   0.94$\pm   0.06(  0.06)$ &   1.10$\pm   0.20(  0.21)$ &   1.00$\pm   0.21(  0.22)$\\
\hline
\multicolumn{8}{c}{synchrotron (at $5.73~\mu$K~K$^{-1}_{408MHz}$)}\\
      11 &   1.32$\pm   0.08$ & -  &   1.31$\pm   0.08(  0.08)$ & -  &   1.04$\pm   0.11(  0.13)$ &   1.11$\pm   0.27(  0.28)$ &   1.01$\pm   0.28(  0.29)$\\
      12 &   1.15$\pm   0.09$ & -  &   1.15$\pm   0.09(  0.09)$ & -  &   1.03$\pm   0.10(  0.10)$ &   1.05$\pm   0.24(  0.26)$ &   1.01$\pm   0.24(  0.26)$\\
      13 &   1.24$\pm   0.13$ & -  &   1.24$\pm   0.12(  0.13)$ & -  &   1.03$\pm   0.15(  0.16)$ &   1.08$\pm   0.32(  0.35)$ &   1.02$\pm   0.32(  0.35)$\\
      14 &   1.22$\pm   0.07$ & -  &   1.22$\pm   0.06(  0.06)$ & -  &   1.04$\pm   0.08(  0.09)$ &   1.08$\pm   0.23(  0.24)$ &   1.00$\pm   0.23(  0.24)$\\
      15 &   0.65$\pm   0.18$ & -  &   0.66$\pm   0.17(  0.19)$ & -  &   1.01$\pm   0.23(  0.25)$ &   0.90$\pm   0.44(  0.45)$ &   1.01$\pm   0.44(  0.45)$\\
\hline
\end{tabular}
\caption{Comparison of resulting fit amplitudes as fraction of true (see text).  The expected foreground residuals after ILC subtraction in the K-band are:  for free-free, assuming $\beta=-2.1$, 0.99\% ;  for dust, added at the \emph{WMAP} best-fit values, approximating a combination of thermal and anomalous dust, 0.94\%;  for synchrotron, assuming $\beta=-3.0$, 1.03\%. $^\dagger$C--C fits using the full covariance matrix. \label{tab:compare_tt_cc} }
\end{table*}

\section{Full fit results}
%
%
\begin{table*}
\begin{tabular}{cccccc}
\hline
Field & K & Ka & Q & V & W \\
\hline
& \multicolumn{5}{c}{\bf free-free}
\\\hline
1 & {\bf   9.8$^{\pm  4.4}$} & {\bf   6.2$^{\pm  4.1}$} & {\bf   6.0$^{\pm  3.9}$} & {\bf   3.4$^{\pm  3.6}$} & {\bf   3.7$^{\pm  3.2}$}\\
2 & {\bf   5.2$^{\pm  2.9}$} & {\bf   1.4$^{\pm  2.7}$} & {\bf   0.3$^{\pm  2.6}$} & {\bf  -0.3$^{\pm  2.5}$} & {\bf   0.7$^{\pm  2.2}$}\\
3 & {\bf   2.3$^{\pm  3.5}$} & {\bf  -1.7$^{\pm  3.2}$} & {\bf  -2.2$^{\pm  3.0}$} & {\bf  -2.5$^{\pm  2.8}$} & {\bf   0.8$^{\pm  2.5}$}\\
4 & {\bf   7.2$^{\pm  2.1}$} & {\bf   1.2$^{\pm  1.1}$} & {\bf   0.9$^{\pm  0.6}$} & {\bf   0.3$^{\pm  0.4}$} & {\bf  -0.2$^{\pm  0.3}$}\\
5 & {\bf  10.1$^{\pm  1.2}$} & {\bf   5.1$^{\pm  1.1}$} & {\bf   3.3$^{\pm  1.0}$} & {\bf   1.6$^{\pm  0.9}$} & {\bf   0.7$^{\pm  0.8}$}\\
6 &  -4.7$^{\pm  9.9}$ & -12.3$^{\pm  7.4}$ & -12.1$^{\pm  6.3}$ &  -5.1$^{\pm  5.8}$ & -10.9$^{\pm  5.2}$\\
7 &   3.5$^{\pm 11.9}$ &   6.8$^{\pm  9.0}$ &  -5.7$^{\pm  7.8}$ &  -0.5$^{\pm  7.2}$ &  -5.2$^{\pm  6.5}$\\
8 & -22.9$^{\pm 16.4}$ &   0.7$^{\pm 13.5}$ &  -5.4$^{\pm 12.3}$ &   2.5$^{\pm 11.3}$ & -13.1$^{\pm 10.2}$\\
9 &   9.2$^{\pm 15.5}$ &  -0.3$^{\pm 11.0}$ &  11.4$^{\pm  9.5}$ &  -1.2$^{\pm  8.7}$ &   0.1$^{\pm  7.9}$\\
10 &  15.3$^{\pm 15.2}$ &  13.6$^{\pm  7.2}$ &   2.7$^{\pm  3.9}$ &   5.7$^{\pm  2.3}$ &   1.1$^{\pm  1.7}$\\
11 & -16.6$^{\pm 19.1}$ & -29.4$^{\pm 13.2}$ &  12.9$^{\pm 11.5}$ &   9.6$^{\pm 10.4}$ &  -2.9$^{\pm  9.3}$\\
12 &   9.0$^{\pm 12.9}$ &   1.6$^{\pm  9.8}$ &  -9.6$^{\pm  8.4}$ &  -7.3$^{\pm  7.7}$ &  -5.2$^{\pm  7.0}$\\
13 &   3.7$^{\pm 14.0}$ &  15.1$^{\pm  7.2}$ &  -3.5$^{\pm  3.7}$ &  -1.5$^{\pm  2.2}$ &  -0.6$^{\pm  1.7}$\\
14 &  -4.4$^{\pm  8.0}$ &  -2.2$^{\pm  5.8}$ &  -1.2$^{\pm  4.9}$ &  -4.9$^{\pm  4.5}$ &   5.8$^{\pm  4.0}$\\
15 &   4.7$^{\pm 17.5}$ &  19.5$^{\pm 12.5}$ &   9.5$^{\pm 10.6}$ &  -9.5$^{\pm  9.7}$ &   2.6$^{\pm  8.7}$\\
\hline
& \multicolumn{5}{c}{\bf dust}
\\\hline
1 &   8.4$^{\pm  5.9}$ &  -0.1$^{\pm  5.1}$ &   2.4$^{\pm  4.3}$ &  -1.0$^{\pm  3.3}$ &  -2.3$^{\pm  2.6}$\\
2 &   6.6$^{\pm  1.6}$ &   1.8$^{\pm  1.3}$ &   2.0$^{\pm  1.2}$ &  -0.8$^{\pm  1.1}$ &   0.3$^{\pm  0.9}$\\
3 &  12.1$^{\pm  5.2}$ &   5.5$^{\pm  4.5}$ &   7.2$^{\pm  3.9}$ &   4.8$^{\pm  3.5}$ &  -0.7$^{\pm  3.1}$\\
4 &   3.9$^{\pm  6.1}$ &   7.3$^{\pm  5.1}$ &   6.4$^{\pm  4.3}$ &   2.5$^{\pm  3.6}$ &  -2.0$^{\pm  3.0}$\\
5 &  12.6$^{\pm  2.1}$ &   2.6$^{\pm  1.8}$ &   2.6$^{\pm  1.6}$ &  -0.3$^{\pm  1.4}$ &  -0.8$^{\pm  1.2}$\\
6 & {\bf   6.7$^{\pm  0.7}$} & {\bf   2.9$^{\pm  0.6}$} & {\bf   2.1$^{\pm  0.6}$} & {\bf   0.8$^{\pm  0.5}$} & {\bf   2.3$^{\pm  0.4}$}\\
7 & {\bf   8.1$^{\pm  3.9}$} & {\bf   2.8$^{\pm  3.5}$} & {\bf   0.2$^{\pm  3.2}$} & {\bf   0.6$^{\pm  2.9}$} & {\bf   0.5$^{\pm  2.6}$}\\
8 & {\bf   8.4$^{\pm  2.5}$} & {\bf   2.6$^{\pm  2.3}$} & {\bf   2.4$^{\pm  2.2}$} & {\bf   2.4$^{\pm  2.0}$} & {\bf   3.1$^{\pm  1.8}$}\\
9 & {\bf   7.3$^{\pm  0.6}$} & {\bf   2.9$^{\pm  0.6}$} & {\bf   1.8$^{\pm  0.5}$} & {\bf   1.4$^{\pm  0.5}$} & {\bf   1.9$^{\pm  0.4}$}\\
10 & {\bf  12.0$^{\pm  1.4}$} & {\bf   4.7$^{\pm  1.2}$} & {\bf   1.8$^{\pm  1.1}$} & {\bf   0.2$^{\pm  1.0}$} & {\bf   1.3$^{\pm  0.9}$}\\
11 &  19.0$^{\pm  7.9}$ &   9.9$^{\pm  6.8}$ &   6.6$^{\pm  6.1}$ &   3.1$^{\pm  5.4}$ &   1.4$^{\pm  4.7}$\\
12 &  15.6$^{\pm  7.5}$ &   5.3$^{\pm  6.6}$ &   4.6$^{\pm  5.8}$ &  -7.1$^{\pm  5.0}$ &  -2.1$^{\pm  4.2}$\\
13 &   3.9$^{\pm  6.4}$ &  -3.7$^{\pm  5.7}$ &   1.6$^{\pm  4.9}$ &  -2.2$^{\pm  3.8}$ &  -1.0$^{\pm  3.1}$\\
14 &   8.3$^{\pm  4.8}$ &   7.6$^{\pm  3.9}$ &   1.8$^{\pm  3.0}$ &   3.0$^{\pm  2.3}$ &  -0.6$^{\pm  1.8}$\\
15 &   8.1$^{\pm  3.3}$ &  -1.0$^{\pm  3.0}$ &  -1.0$^{\pm  2.7}$ &  -1.3$^{\pm  2.5}$ &  -0.5$^{\pm  2.2}$\\
\hline
& \multicolumn{5}{c}{\bf synchrotron}
\\\hline
1 &   3.3$^{\pm  2.4}$ &   0.7$^{\pm  1.0}$ &  -0.7$^{\pm  0.5}$ &   0.0$^{\pm  0.3}$ &  -0.1$^{\pm  0.2}$\\
2 &   8.2$^{\pm  3.5}$ &   0.4$^{\pm  1.5}$ &  -0.9$^{\pm  0.8}$ &   0.8$^{\pm  0.5}$ &   0.7$^{\pm  0.4}$\\
3 &   3.7$^{\pm  2.2}$ &   1.3$^{\pm  0.8}$ &   0.4$^{\pm  0.4}$ &  -0.0$^{\pm  0.2}$ &  -0.1$^{\pm  0.2}$\\
4 &   4.2$^{\pm  3.0}$ &   1.2$^{\pm  1.0}$ &   1.0$^{\pm  0.5}$ &   0.5$^{\pm  0.3}$ &   0.1$^{\pm  0.2}$\\
5 &   4.3$^{\pm  2.2}$ &   1.7$^{\pm  1.1}$ &  -0.2$^{\pm  0.7}$ &  -0.7$^{\pm  0.4}$ &   0.1$^{\pm  0.3}$\\
6 &   2.8$^{\pm  1.3}$ &  -0.3$^{\pm  0.4}$ &   0.1$^{\pm  0.2}$ &  -0.2$^{\pm  0.1}$ &  -0.2$^{\pm  0.1}$\\
7 &   3.8$^{\pm  2.6}$ &   0.6$^{\pm  1.0}$ &  -0.5$^{\pm  0.5}$ &  -0.1$^{\pm  0.4}$ &   0.3$^{\pm  0.3}$\\
8 &   7.2$^{\pm  4.2}$ &   0.6$^{\pm  2.2}$ &   0.0$^{\pm  1.5}$ &  -0.2$^{\pm  1.0}$ &   0.4$^{\pm  0.8}$\\
9 &   0.2$^{\pm  3.3}$ &   0.1$^{\pm  1.6}$ &   2.1$^{\pm  1.0}$ &  -0.5$^{\pm  0.6}$ &  -0.5$^{\pm  0.5}$\\
10 &   1.2$^{\pm  2.1}$ &   0.8$^{\pm  0.9}$ &  -0.4$^{\pm  0.5}$ &   0.6$^{\pm  0.3}$ &   0.1$^{\pm  0.2}$\\
11 & {\bf   4.3$^{\pm  1.8}$} & {\bf  -0.0$^{\pm  1.1}$} & {\bf  -0.6$^{\pm  0.8}$} & {\bf   0.1$^{\pm  0.6}$} & {\bf   0.6$^{\pm  0.5}$}\\
12 & {\bf   2.9$^{\pm  1.6}$} & {\bf   0.6$^{\pm  0.6}$} & {\bf   0.1$^{\pm  0.3}$} & {\bf   0.1$^{\pm  0.1}$} & {\bf   0.1$^{\pm  0.1}$}\\
13 & {\bf   2.8$^{\pm  2.0}$} & {\bf   0.7$^{\pm  0.8}$} & {\bf   0.5$^{\pm  0.4}$} & {\bf   0.0$^{\pm  0.2}$} & {\bf   0.0$^{\pm  0.2}$}\\
14 & {\bf   1.6$^{\pm  1.5}$} & {\bf  -0.4$^{\pm  0.6}$} & {\bf  -0.5$^{\pm  0.3}$} & {\bf  -0.0$^{\pm  0.2}$} & {\bf  -0.0$^{\pm  0.1}$}\\
15 & {\bf   3.0$^{\pm  2.8}$} & {\bf  -0.8$^{\pm  1.5}$} & {\bf  -0.2$^{\pm  0.9}$} & {\bf  -0.4$^{\pm  0.6}$} & {\bf  -0.7$^{\pm  0.5}$}\\
\hline

\end{tabular}\caption{Full fit results for all components in all regions.  The three templates (DDD H$_\alpha$, FDS8 dust, and Haslam 408MHz, plus a constant offset) are fit simultaneously to each band individually. \label{tab:full}} \end{table*}

\label{lastpage}

\end{document}